\documentclass[12pt]{article}

\usepackage{epsfig}
\usepackage{psfrag}
\usepackage{latexsym}
\usepackage[DIV13]{typearea}
\usepackage{amsmath}
\usepackage{amssymb}
\usepackage{amsfonts}
\usepackage{cite}
\usepackage{bbold}
\usepackage[footnotesize]{caption}
\usepackage{graphicx}
\usepackage[center,footnotesize,hang]{subfigure}
\usepackage{url}
\usepackage{color}

\newcommand{\mcal}[1]{\mathcal{#1}}

\newcommand{\al}{\alpha}
\newcommand{\be}{\beta}
\newcommand{\ga}{\gamma}

\newcommand{\de}{\delta}
\newcommand{\De}{\Delta}

\newcommand{\La}{\Lambda}
\newcommand{\om}{\omega}
\newcommand{\sig}{\sigma}
\newcommand{\vphi}{\varphi}

\newcommand{\mev}{\;\text{meV}}

\newcommand{\diag}{\text{diag}}
\newcommand{\hc}{\text{h.c.}}
\newcommand{\unity}{1\hspace{-0.15cm}1}

\newcommand{\mean}[1]{\langle#1\rangle}

\newcommand{\beq}{\begin{equation}}
\newcommand{\eeq}{\end{equation}}
\newcommand{\bac}{\beq\begin{array}}
\newcommand{\eac}{\end{array}\eeq}
\newcommand{\ba}{\begin{array}}
\newcommand{\ea}{\end{array}}
\newcommand{\bea}{\begin{eqnarray}}
\newcommand{\eea}{\end{eqnarray}}

\begin{document}
\begin{titlepage}
\vspace*{-1cm}
\phantom{hep-ph/***}

\hfill{DFPD-09/TH/03}

\hfill{IFIC/09-07}

\vskip 1.5cm
\begin{center}
{\Large\bf  Phenomenological Consequences}

\vskip 0.2 cm
{\Large\bf  of See-Saw in $S_4$ Based Models }
\end{center}
\vskip 0.5  cm
\begin{center}
{\large Federica Bazzocchi}~$^{a)}$\footnote{e-mail address: fbazzoc@few.vu.nl},
{\large Luca Merlo}~$^{b)}$\footnote{e-mail address: merlo@pd.infn.it}
\\
\vskip .2cm
and {\large Stefano Morisi}~$^{c)}$\footnote{e-mail address: morisi@ific.uv.es}
\\
\vskip .2cm
$^{a)}$~Department of Physics and Astronomy, Vrije Universiteit Amsterdam,\\
1081 HV Amsterdam, The Netherlands
\\
\vskip .1cm
$^{b)}$~Dipartimento di Fisica `G.~Galilei', Universit\`a di Padova
\\
INFN, Sezione di Padova, Via Marzolo~8, I-35131 Padua, Italy
\\
\vskip .1cm
$^{c)}$~AHEP Group, Institut de F\'{\i}sica Corpuscular --
  C.S.I.C./Universitat de Val{\`e}ncia \\
  Edificio Institutos de Paterna, Apt 22085, E--46071 Valencia, Spain
\end{center}
\vskip 0.7cm
\begin{abstract}
In \cite{BMM}, it was proposed a flavour model based on the symmetry group $S_4$, managing to describe fermion masses and mixings. The Weinberg operator has been used in order to provide the smallness of the neutrino masses, while a set of scalar fields, getting non-vanishing vacuum expectation values, spontaneously breaks down $S_4$ and provides the Tri-Bimaximal pattern as the lepton mixing matrix. Restricting to this setting, in this paper we analyze possible origins for the effective terms: the type I See-Saw mechanism is the best known approach, but also the type II and III are discussed. The phenomenology related to these models is various and the next future experiments could in principle discriminate among these proposals. Furthermore, we compare our realizations to two relevant $A_4$ based models, also predicting the Tri-Bimaximal lepton mixing, and we find that an analysis on the $0\nu2\be$-decay parameters could distinguish among all these realizations. Furthermore a combined measurement of the effective mass and of the lightest neutrino mass could indicate in the next future which is the preferred flavour symmetry group. The introduction of new physics beyond the Standard Model, like heavy right-handed neutrinos, scalar triplets and fermion triplets, let us investigate on leptogenesis and this provides constraints in the realization of the models.
\end{abstract}
\end{titlepage}
\setcounter{footnote}{0}
\vskip2truecm

%
%

\section{Introduction}

Neutrino experiments managed to measure appearance and disappearance of different flavour neutrinos with good sensitivity \cite{Data} and the so-called atmospheric and solar anomalies can be well explained through the neutrino oscillations. In table \ref{table:OscillationData}, we can read the results of two independent global fits to neutrino oscillation data from \cite{Fogli:Indication} and \cite{Maltoni:Indication}.

\begin{table}[ht]
\begin{center}
\begin{tabular}{lcccc}
\hline
&&&&\\[-2mm]
& \multicolumn{2}{c}{Ref.~\cite{Fogli:Indication}} & \multicolumn{2}{c}{Ref.~\cite{Maltoni:Indication}}\\[2mm]
parameter & best fit (\@$1\sig$) & 3$\sig$-interval & best fit (\@$1\sig)$ & 3$\sig$-interval\\[2mm]
\hline
&&&&\\[-2mm]
$\De m^2_{sol}\:[10^{-5}\mathrm{eV}^2]$
        & $7.67^{+0.16}_{-0.19}$ & $7.14-8.19$
        & $7.65^{+0.23}_{-0.20}$ & $7.05-8.34$\\[2mm]
$|\De m^2_{atm}|\: [10^{-3}\mathrm{eV}^2]$
        & $2.39^{+0.11}_{-0.80}$ & $2.06-2.81$
        & $2.40^{+0.12}_{-0.11}$ & $2.07-2.75$\\[2mm]
$\sin^2\theta_{12}$
        & $0.312^{+0.019}_{-0.018}$ & $0.26-0.37$
        & $0.304^{+0.022}_{-0.016}$ & $0.25-0.37$\\[2mm]
$\sin^2\theta_{23}$
        & $0.466^{+0.073}_{-0.058}$ & $0.331-0.644$
        & $0.50^{+0.07}_{-0.06}$ & $0.36-0.67$\\[2mm]
$\sin^2\theta_{13}$
        & $0.016^{+0.010}_{-0.010}$ & $\leq$ $0.046$
        & $0.010^{+0.016}_{-0.011}$ & $\leq$ $0.056$\\[2mm]
\hline
\end{tabular}
\end{center}
\caption{\label{table:OscillationData} Neutrino oscillation parameters from two independent global fits \cite{Fogli:Indication, Maltoni:Indication}.}
\end{table}

The pattern of the mixings is characterized by two large angles and a small one: the atmospheric angle $\theta_{23}$ is compatible with a maximal value, but the accuracy admits relatively large deviations; the solar angle $\theta_{12}$ is large, but about $5\sigma$ errors far from the maximal value; the reactor angle $\theta_{13}$ has only an upper bound. We underline that there is a tension among the two global fits on the central value of the reactor angle: in \cite{Fogli:Indication} we can read a suggestion for a positive value of $\sin^2\theta_{13}\simeq0.016\pm0.010$ [$1.6\sig$], while in \cite{Maltoni:Indication} the authors find a best fit value consistent with zero within less than $1\sig$. Therefore we need for a direct measurement by the future experiments like DOUBLE CHOOZ \cite{Ardellier:2006mn}, Daya Bay \cite{Wang:2006ca} and MINOS \cite{PereiraeSousa:2005rf} in the $\nu_e$ appearance channel.

A very attractive approximation of the experimental data is provided by the Harrison-Perkins-Scott or Tri-Bimaximal (TB) mixing pattern \cite{TB}
\beq
U_{TB}=\left(
         \begin{array}{ccc}
           \sqrt{2/3} & 1/\sqrt3 & 0 \\
           -1/\sqrt6 & 1/\sqrt3 & -1/\sqrt2 \\
           -1/\sqrt6 & 1/\sqrt3 & +1/\sqrt2 \\
         \end{array}
       \right)\;,
\eeq
which provides the following mixing angles:
\beq
\sin^2\theta_{13}^{TB}=0\qquad\sin^2\theta_{23}^{TB}=1/2\qquad\sin^2\theta_{12}^{TB}=1/3\;.
\eeq
These values are inside the $2\sigma$-ranges of the measured angles and therefore the presence of small perturbations to the TB scheme could improve the agreement with the experimental data: in particular, if a non-vanishing value for the reactor angle $\theta_{13}$ is measured in the future experiments, new ingredients will be necessary in addition to the TB pattern \cite{AFM_Bimax}. On the other hand, the solar angle, $\theta_{12}$, is known with the lowest uncertainty and any deviation for $\theta_{12}^{TB}$ of more than $0.05$ in the wrong direction would rule out this mixing scheme. As a result, without assuming a particular setting for the corrections, the allowed maximal perturbation in the TB angles is about $0.05$.

In the last years there has been lot of efforts in searching for a model which gets the TB patter as the neutrino mixing matrix and a fascinating way seems to be the use of some discrete non-Abelian flavour groups added to the gauge groups of the Standard Model.
There is a series of models based on the symmetry group $A_4$ \cite{TBA4,af:extra,af:modular,afl,afh,bkm,linyin,hmv}, which are particularly attractive: they derive the TB mixing by assuming that the $A_4$ symmetry is realized at a very high energy scale $\La$ and that leptons transform in a non trivial way under this symmetry; afterward the group is spontaneously broken by a set of scalar multiplets $\Phi$, the flavons, whose vacuum expectation values (VEVs) receive a specific alignment. The group $A_4$ is, typically, broken down to two distinct subgroups, one in the neutrino sector and a different one in the charged lepton sector: it is just this breaking chain, distinguishing the two sectors, that produces the TB pattern at the Leading Order (LO). When considering the higher order terms, the TB mixing matrix receives small corrections of order of $\mean{\Phi}/\La<1$ and as a result the reactor angle is no longer vanishing and becomes proportional to $\mean{\Phi}/\La$.

A great difficulty of this kind of models is to describe correctly the quark sector. An interesting solution is to enlarge the symmetry group $A_4$. In particular two non-Abelian discrete groups have been studied: $T'$\cite{fhlm:Tp,Tp} and more recently $S_4$\cite{BMM,S4}. In our last project \cite{BMM}, we propose a supersymmetric flavour model based on the flavour group $S_4\times Z_5\times U(1)_{FN}$, which presents the TB pattern as the lepton mixing matrix, a realistic CKM matrix and correct fermion mass hierarchies. The leading idea is to introduce a set of flavons, that, getting non-vanishing VEVs, break down $S_4$ to its subgroup $Z_2\times Z_2$ in the neutrino sector and to nothing in the charged lepton sector. This breaking chain let us to find the TB scheme at LO as the lepton mixing matrix. Afterwards, the higher order terms introduce small perturbations which deviate the mixing angles from the TB values. On the other hand, the quark mixing matrix comes from a suitable setting of the $S_4$- and $Z_5$-charges.

We use an effective approach to describe neutrinos, indeed their masses are described by the Weinberg operator
\beq
(\ell h_u\ell h_u)\times\ldots
\eeq
where dots represent some flavons. In this paper we are interested in studying possible origins of these effective terms: we study a model based on the flavour group $S_4\times Z_5\times U(1)_{FN}$ by using the See-Saw Mechanism. The simplest approach is the type I See-Saw scheme, but we present also an analysis on the type II and III. Our aim is to find a minimal variation of the previous effective (EF) model, keeping unchanged the scalar and the quark sector: for this reason we do not discuss quark mixing and the mechanism for the vacuum alignment, referring for these aspects to \cite{BMM}. Furthermore we ask to the model to be predictive and therefore we avoid those descriptions which introduce more than two parameters in the structure of the light neutrino mass matrix, like for example models with two different types of See-Saw.

Moreover, we compare our realizations to two relevant $A_4$ based models and we found that an analysis on the $0\nu2\be$-decay parameters could distinguish among all these realizations. A combined measurement of the effective mass and of the lightest neutrino mass could indicate in the next future which is the preferred flavour structure of the neutrino mass matrices.

The introduction of right-handed neutrinos, scalar triplets and fermion triplets could generate a non-vanishing leptogenesis: indeed in all the See-Saw frameworks it is possible to account for CP violating processes. In the latter part of the paper we perform an analysis on the constraints on our proposals from leptogenesis: only the type II See-Saw mechanism is not favoured by this analysis.

The structure of the paper is the following: in section \ref{sec:typeI} we present the model with the type I See-Saw; in section \ref{sec:Phem} there is the analysis of the phenomenological predictions, which we find to be different from those of the EF model; in section \ref{sec:NLO} we discuss the next-to-the-leading order (NLO) corrections and the corresponding deviations to the LO predictions; in section \ref{sec:typeIIandIII} we illustrate the other two possible traditional See-Saw mechanisms, responsible for the lightness of the neutrinos; in section \ref{sec:Leptogenesi} we comment on leptogenesis; finally, section \ref{sec:Conclusions} is dedicated to our conclusions.

%
%

\section{The Model with the Type I See-Saw Mechanism}
\label{sec:typeI}

In this section we illustrate the model in the lepton sector, predicting an exact TB mixing at the LO and a realistic charged lepton mass hierarchy, by the use of the flavour group $G_f$ in addition to the gauge groups of the SM. The complete flavour group is $G_f=S_4\times Z_5\times U(1)_{FN}$, where the three factors play different roles.
$S_4$ is the discrete group given by the permutations of four objects and it has already been studied in literature \cite{S4Old}, but with different aims and different results.\footnote{It has been recently claimed \cite{Lam:S4natural}, through group theoretical arguments, that the minimal flavour symmetry naturally related to the TB mixing is $S_4$. We agree with the conclusions of the group theoretical analysis, but, in our opinion, it must not be considered a constraint for the model realization.} It is composed by 24 elements, divided into $5$ irreducible representations: two singlets, $1_1$ and $1_2$, one doublet, $2$, and two triplets, $3_1$ and $3_2$. We refer to the Appendix A of \cite{BMM} for group details and only recall here the multiplication table for $S_4$:
\beq
\begin{array}{ll}
1_1\otimes R&=R\otimes1_1=R\quad \textrm{with \emph{R} any representation}\\
1_2\otimes1_2&=1_1\\
1_2\otimes2&=2\\
1_2\otimes3_1&=3_2\\
1_2\otimes3_2&=3_1\\
\\
2\otimes2&=1_1\oplus1_2\oplus2\\
2\otimes3_1&=3_1\oplus3_2\\
2\otimes3_2&=3_1\oplus3_2\\
\\
3_1\otimes3_1&=3_2\otimes3_2=1_1\oplus2\oplus3_1\oplus3_2\\
3_1\otimes3_2&=1_2\oplus2\oplus3_1\oplus3_2\;.
\end{array}
\eeq
The spontaneous breaking of $S_4$, with the interplay of the $Z_5$ factor, originates the TB mixing: $S_4$ is broken down to its subgroup $G_\nu=Z_2\times Z_2$ in the neutrino sector\footnote{It is possible to verify it looking to the elements of the group in the appendix A of \cite{BMM}. This is a novelty with respect to the $A_4$ based models, where the main group is broken down to only a $Z_2$ in the neutrino sector.} and to nothing in the charged lepton one. This breaking chain is fundamental in our model, because $G_\nu$ represents the low-energy flavour structure of neutrinos and, in the meantime, breaking $S_4$ down to nothing in the charged lepton sector and setting correctly the $Z_5$ charges, we get a diagonal mass matrix for the charged leptons. As a result, we get that the lepton mixing matrix coincides with the neutrino mixing one, which results to be the TB pattern. Furthermore, the $Z_5$ avoids some unwanted terms, and, with the interplay of the continuous $U(1)_{FN}$ \cite{FN}, is responsible for the hierarchy among the charged lepton masses. In table \ref{table:lepton_transformation}, we can see the lepton sector fields of the model and their transformation properties under $G_f$: with respect to the setting in the EF model, we introduce three right-handed neutrinos which transform as $3_1$ under $S_4$ and we adjust the $Z_5$ charges in a suitable way for our purposes.
\begin{table}[h]
\begin{center}
\begin{tabular}{|c||c|c|c|c|c|c||c||c|c||c|c||c|}
  \hline
  &&&&&&&&&&&& \\[-0,3cm]
  & $\ell$ & $e^c$ & $\mu^c$ & $\tau^c$ & $\nu^c$ & $h_{u,d}$ & $\theta$ & $\psi$ & $\eta$ & $\Delta$ & $\vphi$ & $\xi'$ \\
  &&&&&&&&&&&& \\[-0,3cm]
  \hline
  &&&&&&&&&&&& \\[-0,3cm]
  $S_4$ & $3_1$ & $1_2$ & $1_2$ & $1_1$ & $3_1$ & $1_1$ & $1_1$ & $3_1$ & 2 & $3_1$ & $2$ & $1_2$  \\
  &&&&&&&&&&&& \\[-0,3cm]
  $Z_5$ & $\om^4$ & $1$ & $\om^2$ & $\om^4$ & $\om$ & 1 & 1 & $\om^2$ & $\om^2$ & $\om^3$ & $\om^3$ & 1  \\
  &&&&&&&&&&&& \\[-0,3cm]
  $U(1)_{FN}$ & 0 & 1 & 0 & 0 & 0 & 0 & -1 & 0 & 0 & 0 & 0 & 0  \\
  \hline
  \end{tabular}
\end{center}
\begin{center}
\begin{minipage}[t]{12cm}
\caption{Transformation properties of the matter fields in the lepton sector and of all the flavons of the model. We distinguish the flavon fields on their role and thus we can consider $\psi$ and $\eta$ mainly connected to the charged lepton sector and $\Delta$ and $\vphi$ to the neutrino sector. All these fields together with $\xi'$ are present in the quark sector. The FN field, $\theta$, provides the correct mass hierarchy.}
\label{table:lepton_transformation}
\end{minipage}
\end{center}
\end{table}
We treat the model in a supersymmetric scenario, because the minimization of the scalar potential is simplified, but this is not a constraint from the construction of the model itself.

The superpotential of the model in the lepton sector reads as follows:
\bea
w_\ell\;=&&\sum_{i=1}^{4}\dfrac{\theta}{\La}\dfrac{y_{e,i}}{\La^3}e^c(\ell X_i)'h_d+\dfrac{y_\mu}{\La^2}\mu^c(\ell\psi\eta)'h_d+\dfrac{y_\tau}{\La}\tau^c(\ell\psi)h_d+\hc+\ldots\label{eq:wd:leptons}\\[0.3cm]
w_\nu\;=&&x(\nu^c\ell)h_u+x_d(\nu^c\nu^c\vphi)+x_t(\nu^c\nu^c\De)+\hc+\ldots\label{eq:wd:neutrinos}
\eea
where dots denote higher-order contributions and
\beq
X=\left\{\psi\psi\eta,\;\psi\eta\eta,\;\De\De\xi',\;\De\vphi\xi'\right\}\;,
\label{eq:X}
\eeq
using $(\ldots)$ to refer to the contraction in $1_1$ and $(\ldots)'$ to the contraction in $1_2$. Looking at the table \ref{table:lepton_transformation}, we realize that it is possible to introduce in $w_\nu$ other two terms, $\ell h_u\ell h_u\vphi/\Lambda^2$ and $\ell h_u\ell h_u\Delta/\Lambda^2$: we are assuming that $\nu^c$ are the only origins of the light neutrino masses and therefore we can omit those operators, which would account for other possible sources. It will be clear in a while that these contributions are suppressed with respect the type I terms.

We underline that the first contributions containing $e^c$ would be
\beq
\dfrac{\theta}{\La}\dfrac{y'_{e,1}}{\La^2}e^c(\ell\De\De)'h_d+\dfrac{\theta}{\La}\dfrac{y'_{e,2}}{\La^2}e^c(\ell\De\vphi)'h_d\;,
\eeq
which would dominate with respect to the terms in eq.(\ref{eq:wd:leptons}). However an explicit computation will show that these two terms are vanishing, once we assume that the flavons get these specific VEVs:
\bac{rclrcl}
\mean{\psi}&=&\left(
             \begin{array}{c}
               0 \\
               1 \\
               0 \\
             \end{array}
           \right)v_\psi&
\mean{\eta}&=&\left(
             \begin{array}{c}
               0 \\
               1 \\
             \end{array}
           \right)v_\eta\\
\\
\mean{\Delta}&=&\left(
             \begin{array}{c}
               1 \\
               1 \\
               1 \\
             \end{array}
           \right)v_\Delta&
\mean{\vphi}&=&\left(
             \begin{array}{c}
               1 \\
               1 \\
             \end{array}
           \right)v_\vphi\\
\\
\mean{\xi'}&=&v_{\xi'}&
\mean{\theta}&=&v_\theta
\label{vev:allleptons}
\eac
with
\beq
v_\De^2=-\dfrac{g_3}{3g_2}v_\vphi^2\quad\quad v_\psi=-\dfrac{f_2}{2f_1}v_\eta\quad\quad v_{\xi'}=\dfrac{h_1}{M_{\xi'}}v_\eta v_\vphi\quad\quad|\theta|^2=M^2_{FI}/g_{FN}
\eeq
and $v_\vphi$ and $v_\eta$ undetermined. The factors $g_i$, $f_i$, $h_1$, $M_{\xi'}$, $M_{FI}$ and $g_{FN}$ are masses and coupling constants which appear in the superpotential containing the flavons and it is possible to show that this particular VEV alignment\footnote{Some attempts in which the VEV alignment problem in not present can be found in \cite{NoAlignment}.} is a natural solution of the scalar potential and that it is stable under the higher order corrections (the complete discussion can be found in \cite{BMM}). All the VEVs are of the same order of magnitude and for this reason we will parameterize the ratio $VEV/\La$ by the parameter $u$. The only VEV which originates a different mechanism with respect to the others is $v_\theta$ and we indicate the ratio $v_\theta/\La$ by the parameter $t$. With a simple analysis on the hierarchies of the fermion masses we get \cite{BMM} that $u$ and $t$ belong to a well determined range of values:
\beq
0.01<u,t<0.05\,.
\label{vev:uet}
\eeq

With this setting, in the basis of canonical kinetic terms, the mass matrix for the charged leptons is ($m_\ell\sim R^cL$, $m_\nu\sim L^TL$)
\beq
m_\ell=\left(
         \begin{array}{ccc}
           y_e^{(1)} u^2t & y_e^{(2)} u^2t & y_e^{(3)} u^2t \\
           0 & y_\mu u & 0 \\
           0 & 0 & y_\tau \\
         \end{array}
       \right)uv_d
\label{matrices:l}
\eeq
where the $y_e^{(i)}$ are the result of all the different contributions of the $y_{e,i}$.
The neutrino mass matrices are
\beq
m_\nu^D=\left(
          \begin{array}{ccc}
            1 & 0 & 0 \\
            0 & 0 & 1 \\
            0 & 1 & 0 \\
          \end{array}
        \right)xv_u\qquad\qquad
M_N=\left(
          \begin{array}{ccc}
            2c & b-c & b-c \\
            b-c & b+2c & -c \\
            b-c & -c & b+2c \\
          \end{array}
        \right)
\label{matrices:nu}
\eeq
where $b=2x_dv_\vphi$ and $c=2x_tv_\Delta$. The heavy neutrino mass matrix $M_N$ is $2 \leftrightarrow 3$ invariant and it satisfies the relation $(M_N)_{11}+(M_N)_{13}=(M_N)_{22}+(M_N)_{23}$, which is the so-called magic symmetry. It is a general result that a mass matrix with these properties can be exactly diagonalized by the TB mixing and therefore
\beq
M_N^{diag}=(3c-b,\;2b,\;3c+b)\;.
\eeq
Integrating out the heavy degrees of freedom, we get the light neutrino mass matrix, which is given by the following relation,
\beq
m_\nu=-(m_\nu^D)^TM_N^{-1}m_\nu^D\;.
\label{matrices:LHnu}
\eeq
The explicit form of $m_\nu$ is not simple, but it is still diagonalized by the TB pattern. The light neutrino mass eigenvalues are simply the inverse of the heavy neutrino ones, a part from a minus sign and the global factors from $m_\nu^D$:
\beq
m_{\nu_1}=-\dfrac{x^2v_u^2}{3c-b}\qquad m_{\nu_2}=-\dfrac{x^2v_u^2}{2b}\qquad m_{\nu_3}=-\dfrac{x^2v_u^2}{3c+b}\;.
\label{LNMassEigen}
\eeq
It is possible to get a naive esteem on the mass of the heavy right-handed neutrinos: indeed from eq.(\ref{matrices:LHnu}) we can write
\beq
m_\nu\sim\dfrac{x^2v_u^2}{M_N}
\label{stimaMN}
\eeq
and, taking $v_u=174$ GeV and $x$ of $\mcal{O}(1)$ and considering $|\De m_{atm}^2|^{1/2}$ has the typical light neutrino mass scale, we get that the typical mass scale for the heavy neutrinos is $5\times10^{14}$ GeV.

It is possible now to underline that the type I terms give a larger contribution to the neutrino masses than the effective ones, $\ell h_u\ell h_u\vphi/\Lambda^2$ and $\ell h_u\ell h_u\Delta/\Lambda^2$. When the Higgs fields and the flavons get VEV, then the contributions to the light neutrino masses are $m_\nu^\mathrm{eff}\sim u\, v_u^2/\Lambda$. On the other hand, looking at eq.(\ref{stimaMN}) we get that $m_\nu^\mathrm{type\,I}\sim (1/u)\,v_u^2/\Lambda$. Since $u<1$, then we conclude that $m_\nu^\mathrm{eff}<m_\nu^\mathrm{type\,I}$.

In the next section we analyze the phenomenology of this model and we compare it with the results of the EF model.

%
%

\section{Phenomenological Analysis}
\label{sec:Phem}

Considering the left handed neutrino mass eigenvalues of eq.(\ref{LNMassEigen}) we can write the neutrino oscillation parameters $\De m_{atm}^2$ and $\De m_{sol}^2$ as follows:
\begin{eqnarray}
&\Delta m_{atm}^2&=|m_{\nu_3}|^2-|m_{\nu_1}|^2=-\dfrac{12|b||c|\cos \zeta}{|b|^4+81|c|^4-18|b|^2|c|^2\cos2\zeta}|x|^4v_u^4\\[0.3cm]
&\Delta m_{sol}^2&=|m_{\nu_2}|^2-|m_{\nu_1}|^2=\left(\dfrac{1}{4|b|^2}-\dfrac{1}{|b|^2+9|c|^2-6|b||c|\cos\zeta}\right)|x|^4v_u^4
\end{eqnarray}
where $\zeta$ is the relative phase between $b$ and $c$. This phase is related by a non-trivial relation to the Majorana CP phase $\al_{21}$, which is defined as follows
\beq
U_\nu=U_{TB}\cdot\diag\left(1,e^{i\frac{\al_{21}}{2}},e^{i\frac{\al_{31}}{2}}\right)\;.
\eeq
We can express $|b|$ and $|c|$ as functions of $\Delta m_{atm}^2$, $r\equiv\Delta m_{sol}^2/|\Delta m_{atm}^2|$ and $\zeta$ and as a result we get constraints on the type of the neutrino spectrum, on the value of the lightest neutrino mass and on the effective $0\nu2\be$-decay mass, $|m_{ee}|$, directly from the experimental data \cite{Hirsch}.

\begin{figure}[h!]
 \centering
 \subfigure[Present proposal.]
   {\includegraphics[width=8cm]{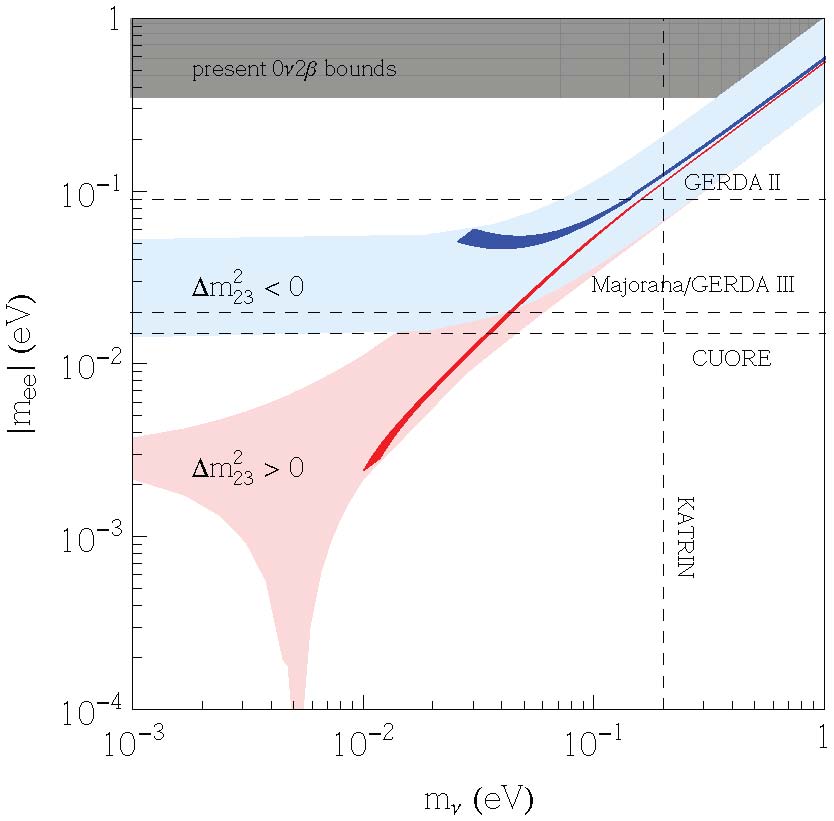}}
 \subfigure[Analysis from \cite{BMM}.]
   {\includegraphics[width=8cm]{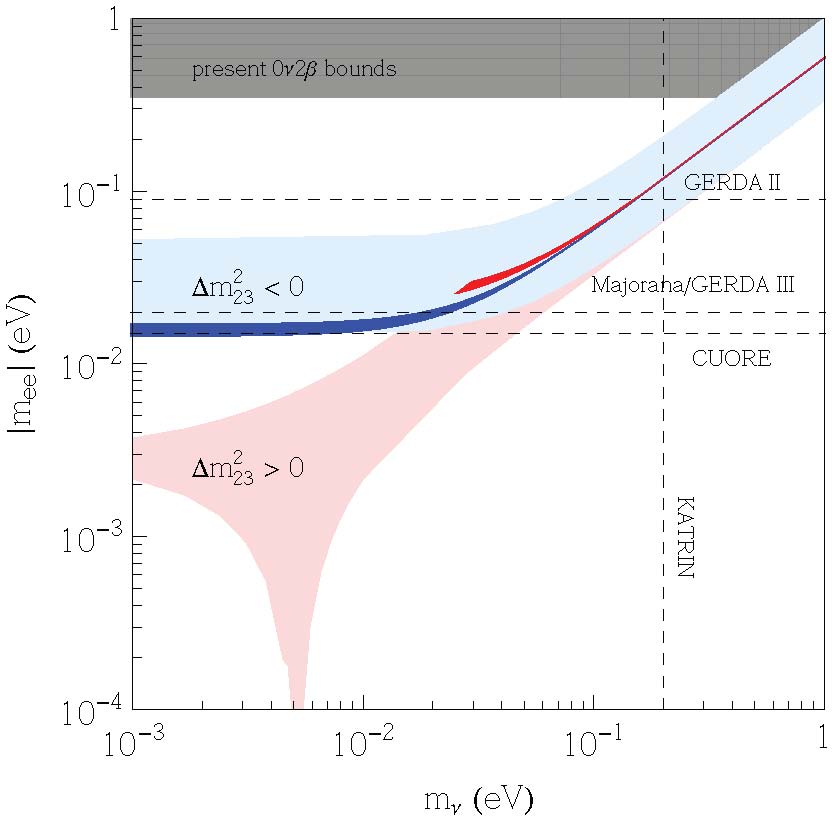}}
 \caption{In both figures, $|m_{ee}|$ as a function of the lightest neutrino mass is plotted. The red corresponds to the NH case and the blue to the IH one. The light colored bands represent the possible regions considering only the exact TB pattern, while the dark colored ones are the predictions of the specific models: on the left the present proposal and on the right the analysis in \cite{BMM}. The present bound from the Heidelberg-Moscow experiment is shown in dark gray and the future sensitivity of CUORE ($\sim15$ meV), Majorana and GERDA III ($\sim20$ meV), and GERDA II ($\sim90$ meV) experiments are represented by the horizontal dashed lines, while the future sensitivity of $0.2$ eV of KATRIN experiment is shown by the vertical dashed line.}
 \label{fig1}
\end{figure}

In figure \ref{fig1}, we plot $|m_{ee}|$ as a function of the lightest neutrino mass: we compare the scenarios of the present proposal with the type 1 See-Saw mechanism (SS1) and our previous analysis \cite{BMM} with the effective approach. The light colored bands correspond to the different types of neutrino mass spectrum for the allowed values of $|m_{ee}|$ for the exact TB pattern, while the dark colored ones to the predictions of the two models: the Normal Hierarchy (NH), in red, and the Inverted Hierarchy (IH), in blue.  The constraints which have been imposed to draw all the regions are the experimental values at $3\sig$ for $\De m^2_{atm}$ and for the ratio $r=0.032^{+0.006}_{-0.005}$ \cite{Maltoni:Indication}. In dark gray, we plot the present bound from the Heidelberg-Moscow \cite{HM} experiment on $|m_{ee}|$. The vertical dashed line corresponds to the future sensitivity on the lightest neutrino mass of $0.2$ eV from the KATRIN experiment\cite{katrin} and the horizontal ones to the future sensitivity of some $0\nu2\be$-decay experiments, that are $15$ meV, $20$ meV and $90$ meV, respectively of CUORE \cite{cuore}, Majorana \cite{majorana}/GERDA III\cite{gerda} and GERDA II experiments.

In the model with See-Saw mechanism, figure \ref{fig1}(a), we cannot refer to the IH, because the IH region falls in the Quasi Degenerate (QD) case band and the two hierarchies will not be distinguished any more by experiments (in the following, when we refer to the IH for the SS1 model, we will have in mind this aspect). Furthermore, the QD spectrum is subjected to a moderate fine-tune in order to fit the value of $r$: indeed the mass differences have to be in well determined relations. From this point of view the NH seems to be the favoured neutrino spectrum. The NH and the QD, however, can be distinguished in the near future experiments: an absence of any signal linked to some lepton number violating processes would fix an experimental upper bound on $|m_{ee}|$ which completely would rule out the QD and allow only the NH.

It is possible to repeat the discussion on the IH of the SS1 model for the NH in the EF model: indeed the NH region falls in the QD area. Therefore in the model with the effective approach, figure \ref{fig1}(b), only the IH and the QD can be explained, even if the QD spectrum is subjected to a small fine-tuning in order to account for $r$. Finally we can conclude that the favourite spectrum is the inverted hierarchical one. Differently from the SS1 model, the absence of any measured lepton number violation in the next experiments would almost completely rule out this model. On the other hand, if some lepton number violating process is measured, it will be important also the information on the mass of the lightest neutrino: from figure \ref{fig1}(b), a value for $m_\nu$ less than about $0.02$ eV will imply that only the IH is recovered. Unfortunately, the experimental sensitivity is far from this value, indeed the lowest one is from KATRIN experiment and it will be $0.2$ eV.

\begin{figure}[h!]
 \centering
 \subfigure[Present proposal.]
   {\includegraphics[width=8cm]{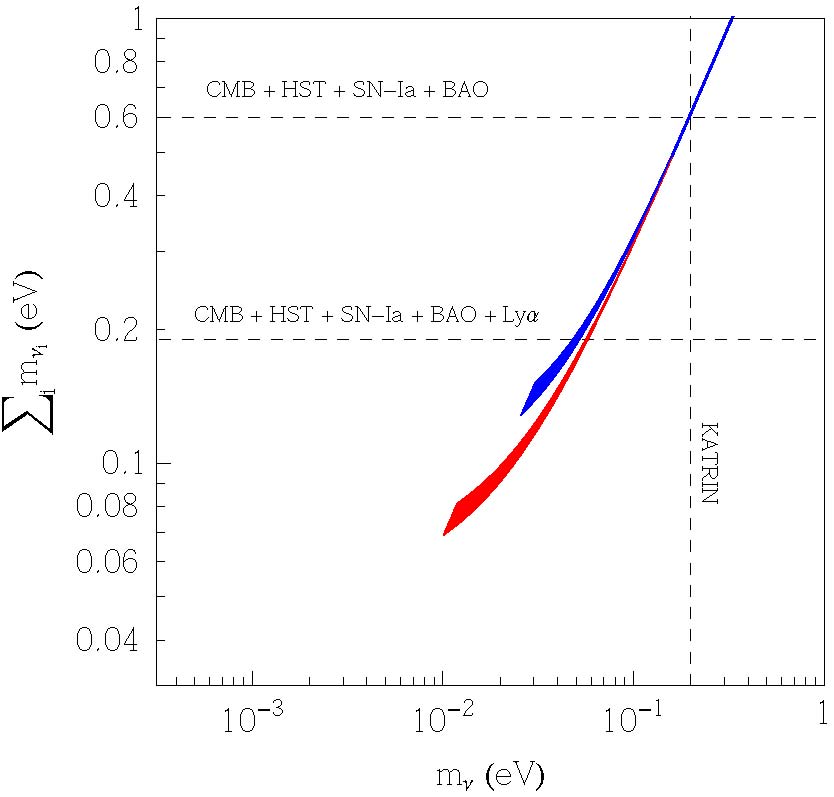}}
 \subfigure[Analysis from \cite{BMM}.]
   {\includegraphics[width=8cm]{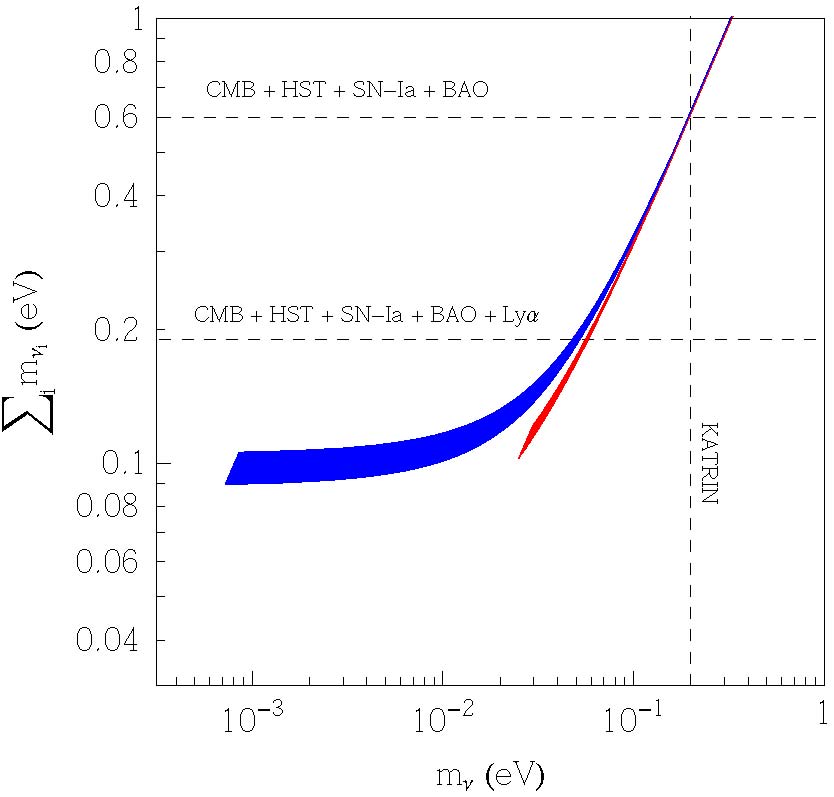}}
 \caption{In both figures, the sum of the neutrino masses as a function of the lightest mass. In red the NH case and in blue the IH one. On the left the present proposal and on the right the analysis in \cite{BMM}. The vertical dashed line represents the future sensitivity of $0.2$ eV from KATRIN experiment and the horizontal ones refer to the cosmological bounds (see the text for more details).}
 \label{fig2}
\end{figure}

In figure \ref{fig2}, we continue the comparison between the two models, plotting the sum of the neutrino masses as a function of the lightest neutrino mass. The vertical dashed line refers to the future sensitivity of KATRIN experiment, while the horizontal ones to the cosmological bounds \cite{Cosmology}. There are typically five representative combinations of the cosmological data, which lead to increasingly stronger upper bounds on the sum of the neutrino masses: we are showing the two strongest ones. The first one at $0.60$ eV corresponds to the combination of the Cosmic Microwave Background (CMB) anisotropy data (from WMAP~5y \cite{WMAP2}, Arcminute Cosmology Bolometer Array Receiver (ACBAR) \cite{acbar07}, Very Small Array (VSA) \cite {vsa}, Cosmic Background Imager (CBI) \cite{cbi} and BOOMERANG \cite{boom03} experiments) plus the large-scale structure (LSS) information on galaxy clustering (from the Luminous Red Galaxies Sloan Digital Sky Survey (SDSS) \cite{Tegmark}) plus the Hubble Space Telescope (HST) plus the luminosity
distance SN-Ia data of \cite{astier} and finally plus the BAO data from \cite{bao}. The second line at $0.19$ eV corresponds to all the previous data combined to the small scale primordial spectrum from Lyman-alpha (Ly$\alpha$) forest clouds \cite{Ly1}.

Looking at the plots, we see that the cosmological bounds cannot discriminate among the two models, the SS1 in figure \ref{fig2}(a) and the EF in figure \ref{fig2}(b). If some progress is done in order to lower these bounds, however, it could be possible to discriminate from the NH and the IH in the SS1 model.

\begin{figure}[h!]
 \centering
 \subfigure[Present proposal.]
   {\includegraphics[width=8cm]{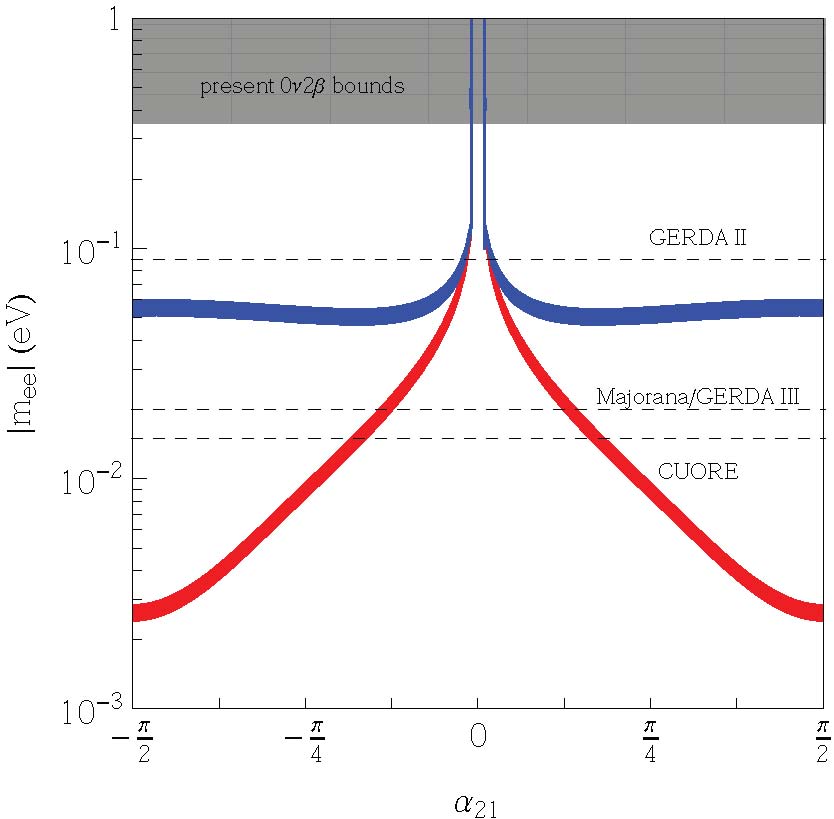}}
 \subfigure[Analysis from \cite{BMM}.]
   {\includegraphics[width=8cm]{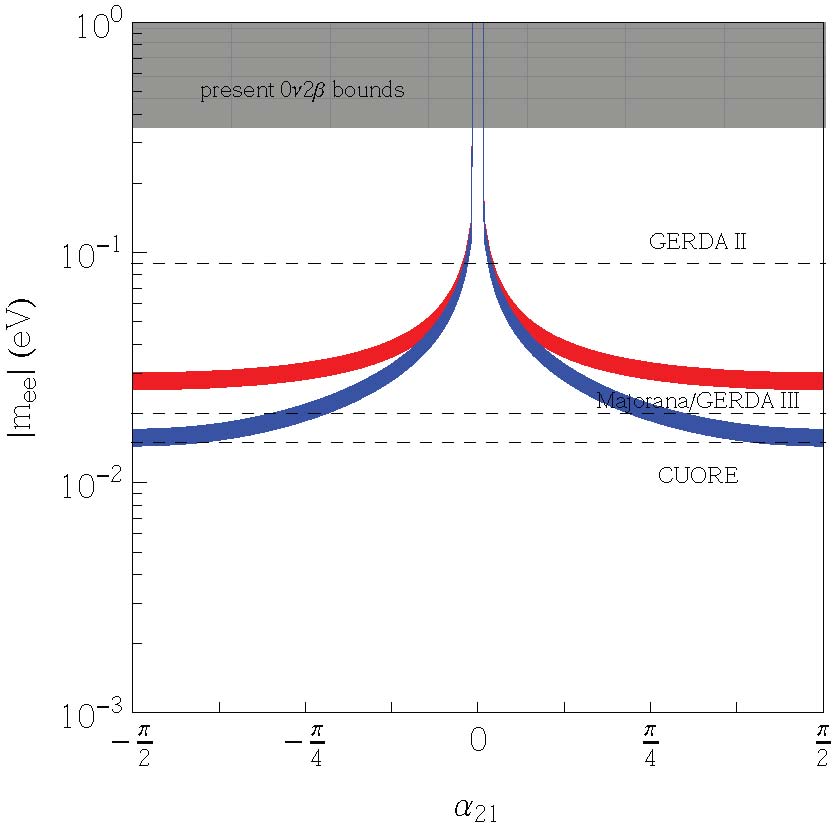}}
 \caption{In both figures, $|m_{ee}|$ as a function of the physical Majorana phase $\alpha_{21}$. In red the NH case and in blue the IH one. On the left the present proposal and on the right the analysis in \cite{BMM}. The present bound from the Heidelberg-Moscow experiment is shown in dark gray and the future sensitivity of CUORE ($\sim15$ meV), Majorana and GERDA III ($\sim20$ meV), and GERDA II ($\sim90$ meV) experiments are dashed.}
 \label{fig3}
\end{figure}

Finally in figure \ref{fig3}, we plot $|m_{ee}|$ as a function of the physical Majorana phase. From the present bounds, the CP-conserving case, $\alpha_{21}=0$, is completely excluded for both the models. Moreover, these plots confirm the results from figure \ref{fig1}: the next experiments on $|m_{ee}|$ could completely rule out the effective model and the IH of the model with See-Saw, in the case of an absence of any measured lepton number violating process.

\begin{table}[h]
\centering
\begin{tabular}{|c|c|c|c|}
  \hline
  &&&\\[-3mm]
  & $|m_{ee}|$ (meV) & lightest $m_\nu$ (meV) & $\sum_i|m_{\nu i}|$ (meV)\\[3mm]
  \hline
  &&&\\[-3mm]
  See-Saw model & 2.3 & 9.5 & 67.3 \\[3mm]
  Effective model & 14.4 & 0.72 & 89.4 \\[3mm]
  \hline
\end{tabular}
\caption{\label{table:comparison}Lower bounds for $|m_{ee}|$, the lightest neutrino mass and the sum of the neutrino masses. The first row refers to the model with the See-Saw mechanism, while the second one to the proposal in \cite{BMM}.}
\end{table}

In table \ref{table:comparison}, we show the lower bounds for $|m_{ee}|$, the lightest neutrino mass and the sum of the neutrino masses, for both the models: we remember that, only the NH can be explained in the See-Saw model, while only the IH does in the effective approach, without any fine-tuning.

All these quantities can help in distinguishing our models from the others, which present the TB pattern; furthermore the predictions for the lower bound on $|m_{ee}|$ are close to the future experimental sensitivity, which are expected to reach the values of $0.090\textrm{ eV}$\cite{gerda} (GERDA II), $0.020\textrm{ eV}$\cite{majorana} (Majorana/GERDA III), $0.050\textrm{ eV}$\cite{supernemo} (SuperNEMO), $0.015\textrm{ eV}$\cite{cuore} (CUORE) and $0.024\textrm{ eV}$\cite{exo} (EXO).

\subsection{Comparison with other Models}
In this part we compare our results with two phenomenological interesting models, which still predict the TB pattern and present a neutrino mass matrix with only two parameters: with these requirements we select a quite little set of models, that can be directly compared with our realizations.

\subsubsection{The Altarelli-Feruglio Model}
The first proposal we consider is the Altarelli-Feruglio (AF) model. We do not enter in the details of the model building aspects, for which we refer to the original papers \cite{af:extra,af:modular,afl}, but we only briefly resume the main aspects. The AF model is based on the symmetry group $A_4\times Z_3\times U(1)_{FN}$, under which the leptons transform in a non-trivial way. Some flavons are added to the particle spectrum and, when they get non-vanishing VEVs, they spontaneously break $A_4$ down to its subgroup $Z_2$ in the neutrino sector and to its subgroup $Z_3$ in the charged lepton sector. As a result, the following neutrino mass matrix is recovered
\beq
m_\nu=\left(
    \begin{array}{ccc}
    Y+2Z& -Z& -Z\\
    -Z& 2Z& Y-Z\\
    -Z& Y-Z& 2Z
    \end{array}\right)\frac{v_u^2}{\Lambda}\;,
\eeq
which is diagonalized by the TB pattern. On the other hand, the charged leptons are diagonal and then the lepton mixing matrix coincides with the neutrino one, that is the TB pattern. This LO result is subsequently corrected by the higher order terms and thus it predicts a non-vanishing reactor angle. The smallness of the neutrino masses is explained through the Weinberg operator, dealing with the effective approach. Furthermore, if heavy right-handed neutrinos are added, with non trivial transformation properties under the flavour symmetry group, the model accounts for the type I See-Saw mechanism.

Before discussing the phenomenology related to the AF model, it seems of worth to mention the realization with the discrete symmetry group $T'$: in \cite{fhlm:Tp}, the authors reproduce the flavour structure of the neutrino mass matrix and the results of the AF model in the neutrino sector, introducing a description for the quark sector. As a result the following analysis on the AF model is valid also for the $T'$ model of \cite{fhlm:Tp}.

\begin{figure}[h!]
 \centering
 \subfigure[AF model --- See-Saw case.]
   {\includegraphics[width=8cm]{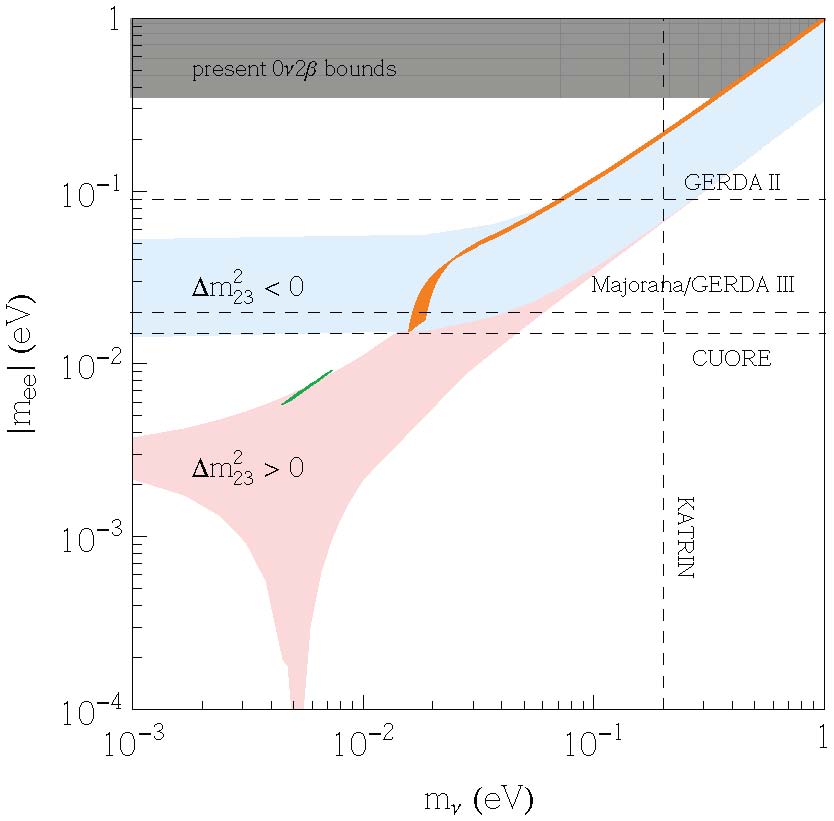}}
 \subfigure[AF model --- Effective case.]
   {\includegraphics[width=8cm]{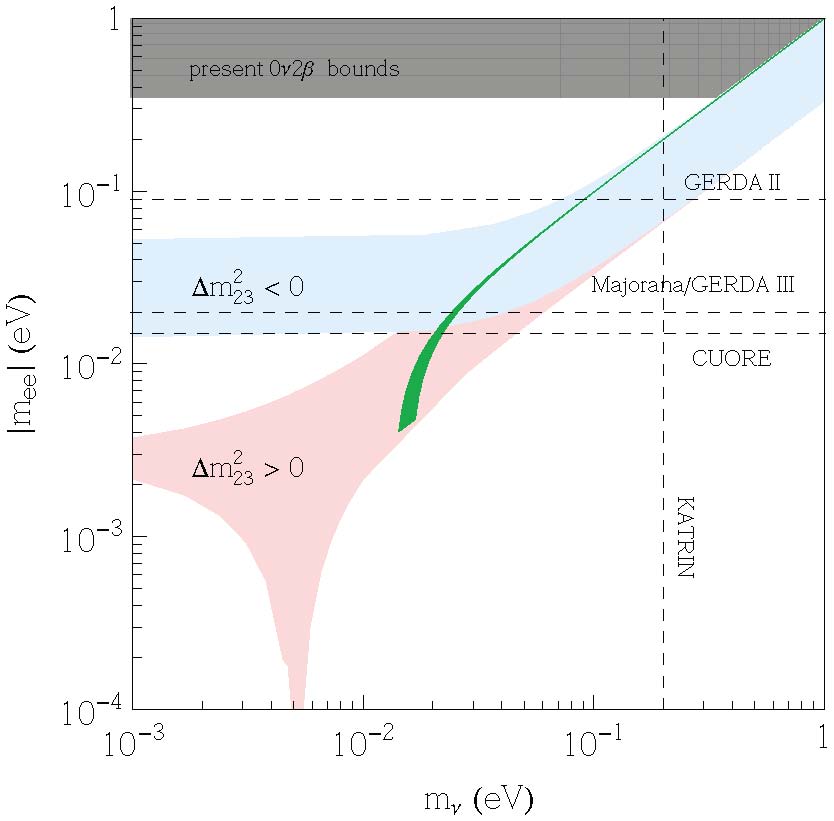}}
 \caption{In both figures, $|m_{ee}|$ as a function of the lightest neutrino mass is plotted. The light colored regions represent the allowed areas considering only the exact TB pattern: in red the NH and in blue the IH. The dark colored bands are the predictions of the AF model: on the left the case with the See-Saw scheme and on the right the effective case. In green is represented the NH and in orange the IH. The experimental bounds are the same as in figure \ref{fig1}.}
 \label{fig4}
\end{figure}

In figure \ref{fig4}, we show the predictions for the effective mass $|m_{ee}|$ as a function of the lightest neutrino mass in both the versions of the model: in figure \ref{fig4}(a) the realizations with the See-Saw scheme and in figure \ref{fig4}(b) with the effective approach. The constraints which have been imposed to draw all the regions are the experimental values at $3\sig$ for $\De m^2_{atm}$ and for the ratio $r$.

As we can see, in the See-Saw framework both the hierarchies can be accommodated, even if the IH covers a larger band of the space of the parameters: as reported in \cite{af:modular}, in order to explain the ratio $r$, the parameters which define the neutrino mass eigenvalues have to undergo to a small fine-tuning and as a result the space of the parameters contracts and in the case of the NH the allowed range for $|m_{ee}|$ is small. We can see at this result as a very precise prediction of the model and indeed the ranges for $|m_{ee}|$, the lightest neutrino mass and the sum of the masses in the case of the NH are
\bea
&5.8\mev<|m_{ee}|<9.2\mev\\[3mm]
&4.4\mev<m_{\nu_1}<7.3\mev\\[3mm]
&57.9\mev<\sum_im_{\nu i}<73.6\mev\;.
\eea
With respect to all the other models discussed in the paper, only this one presents upper bounds for the three observables just recalled: it is a very attractive feature which characterizes the AF realization with the type I See-Saw mechanism.\\

The situation changes in the effective scenario, where only the NH and the QD spectrum can be accommodated. The two realizations, with and without right-handed neutrinos, will be difficult to distinguish one from each other: indeed only precise combined measurements of $|m_{ee}|$ and the lightest neutrino mass could discriminate between them.

\begin{table}[h]
\centering
\begin{tabular}{|c|c|c|c|}
  \hline
  &&&\\[-3mm]
  & $|m_{ee}|$ (meV) & lightest $m_\nu$ (meV) & $\sum_i|m_{\nu i}|$ (meV)\\[3mm]
  \hline
  &&&\\[-3mm]
  See-Saw model -- NH case & 5.8 -- 9.2 & 4.4 -- 7.3 & 57.9 -- 73.6 \\[3mm]
  See-Saw model -- IH case & 15.2 & 15.7 & 109.6 \\[3mm]
  \hline
  &&&\\[-3mm]
  Effective model & 3.7 & 13.8 & 77.2 \\[3mm]
  \hline
\end{tabular}
\caption{\label{table:comparisonAF}$|m_{ee}|$, the lightest neutrino mass and the sum of the neutrino masses for the AF model. The first row refers to the allowed ranges for these observables for the NH in the See-Saw case, while the second and third rows to their lower bounds for the IH in the See-Saw case and for the realization with the effective approach.}
\end{table}

In table \ref{table:comparisonAF}, we report the allowed ranges for $|m_{ee}|$, the lightest neutrino mass and the sum of the masses for the AF model with the See-Saw mechanism in the NH case and the lower bounds for these observables in the realization with the See-Saw mechanism in the IH case and in that one with the effective approach.

\begin{figure}[h!]
 \centering
 \subfigure[Comparison between the SS1 and the AF models.]
   {\includegraphics[width=8cm]{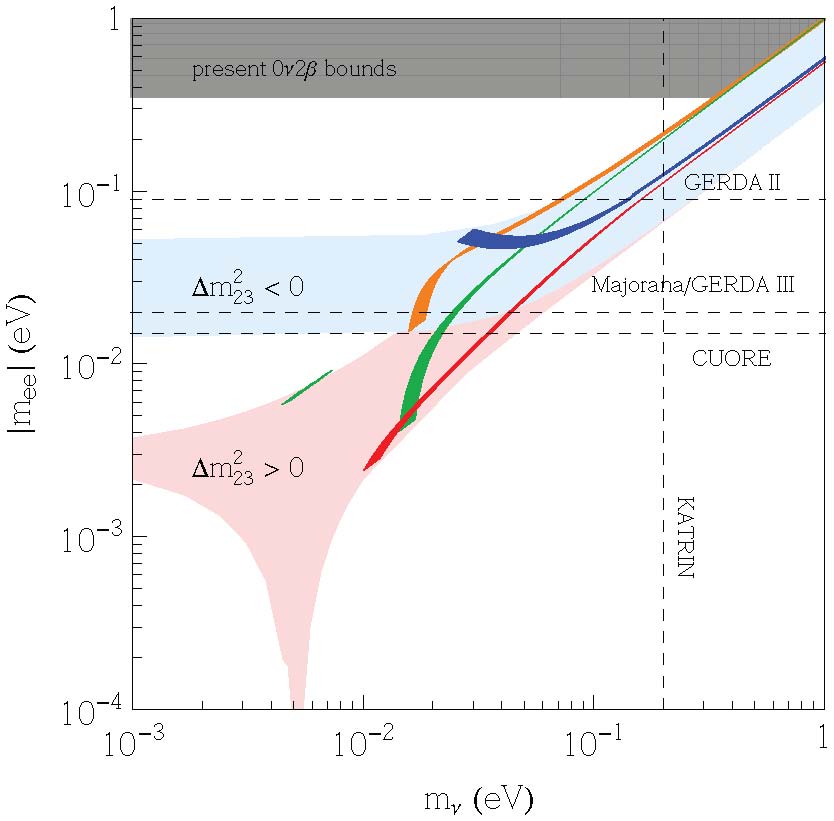}}
 \subfigure[Comparison between the EF and the AF models.]
   {\includegraphics[width=8cm]{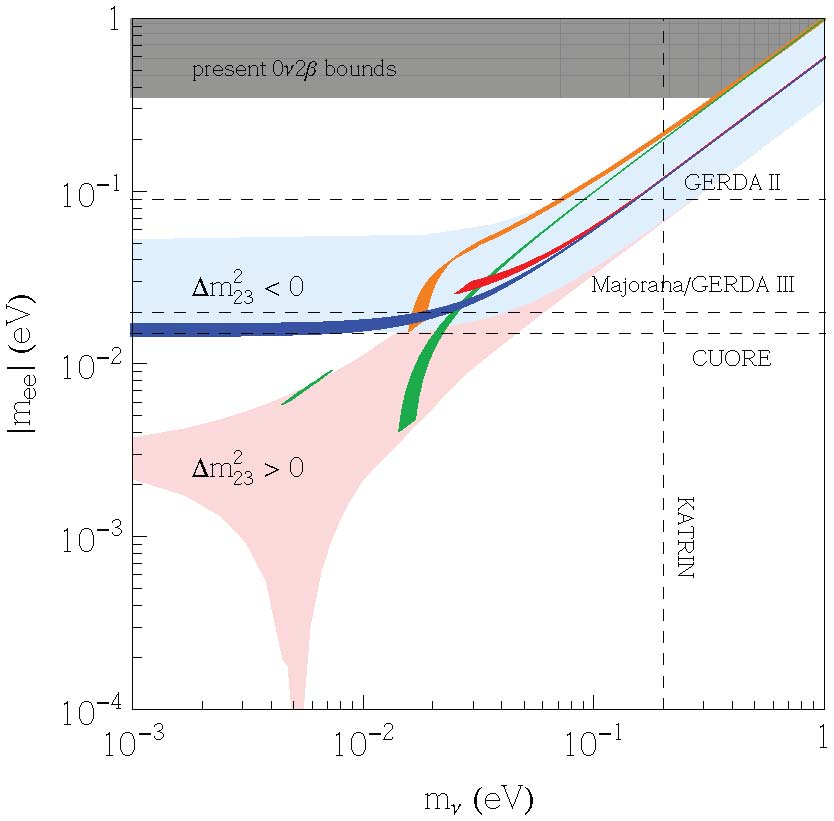}}
 \caption{In both figures, $|m_{ee}|$ as a function of the lightest neutrino mass is plotted. We compare our proposals and the AF model: on the left we plot the SS1 model and on the right the EF one. In red (green) and blue (orange) we plot the NH and the IH, respectively, of our proposals (AF model). The experimental bounds are the same as in figure \ref{fig1}.}
 \label{fig5}
\end{figure}

We can compare our proposals with the possible scenarios of the AF model. In figure \ref{fig5}, we overlap the plots already shown in figures \ref{fig1} and \ref{fig4}: in figure \ref{fig5}(a) there is the SS1 realization, while in figure \ref{fig5}(b) the EF one. The models cover essentially the same part of the space of the parameters and therefore it is not simple to distinguish among them. However there is an extremely interesting aspect which can discriminate among the AF model and the $S_4$ based proposals: in the region of the QD spectrum, the models dispose along two distinct and parallel lines. This different behaviour has to be addressed to the Majorana phase $\alpha_{21}$: in order to have degenerate neutrino masses in the AF model it has to be required $\alpha_{21}=0$, while in our proposals it has to be $\alpha_{21}=\pi$. The origin of this splitting is the flavour structure of the neutrino mass matrices. In the next future experiments, a combined measure of $|m_{ee}|$ and of the lightest neutrino mass can definitively rule out one of the two descriptions in the QD region.\\

\subsubsection{The Niemeyer Model}

We move now to another analysis based on the $A_4$ symmetry group and recently presented in \cite{hmv}.
In this paper, the authors deal only with the type I See-Saw mechanism where the Dirac neutrino mass matrix, $m_\nu^D$, and the right-handed Majorana neutrino one, $M_N$, are the following:
\beq
m_\nu^D\sim\left(
          \begin{array}{ccc}
            Y & 0 & 0 \\
            0 & Y & Z \\
            0 & Z & Y \\
          \end{array}
        \right)\qquad\qquad
M_N\sim\left(
         \begin{array}{ccc}
           1 & 0 & 0 \\
           0 & 1 & 0 \\
           0 & 0 & 1 \\
         \end{array}
       \right)\;,
\eeq
from which the light neutrino mass matrix is given through the traditional relation
\beq
m_\nu=-(m_\nu^D)^TM_N^{-1}m_\nu^D\;.
\eeq
The charged leptons are initially not diagonal, but they can be rotated in the diagonal form by the use of the unitary matrix $U_\om$,
\beq
U_\om=\dfrac{1}{\sqrt3}\left(
                         \begin{array}{ccc}
                           1 & 1 & 1 \\
                           1 & \om & \om^2 \\
                           1 & \om^2 & \om \\
                         \end{array}
                       \right)\;.
\eeq
Moving to this basis, $U_\om$ has to be applied to $m_\nu$ and the final result is a light neutrino mass matrix which is diagonalized by the TB pattern.

They present the study of the phenomenology related to this particular structure of the neutrino mass matrices, which results to have interesting predictions with respect to the other present realizations. Furthermore they indicate two alternative settings, with different contents of fields in the Higgs sector, which can reproduce such matrices.

\begin{figure}[h!]
 \centering
 \subfigure[The Niemeyer model.]
   {\includegraphics[width=8cm]{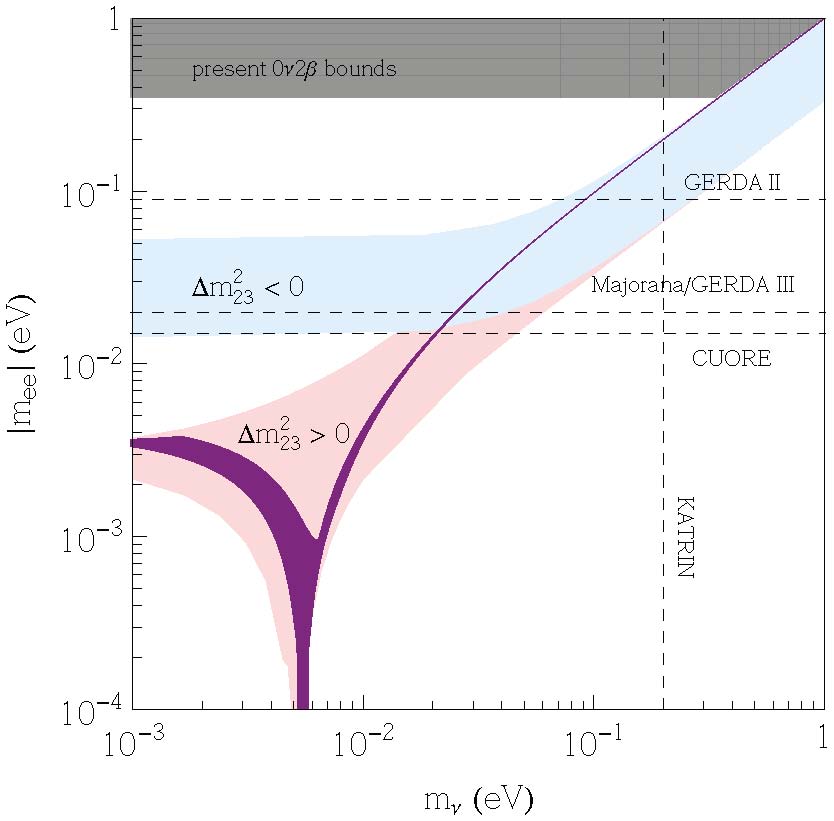}}
 \subfigure[The ``Niemeyer" plot.]
   {\includegraphics[width=8cm]{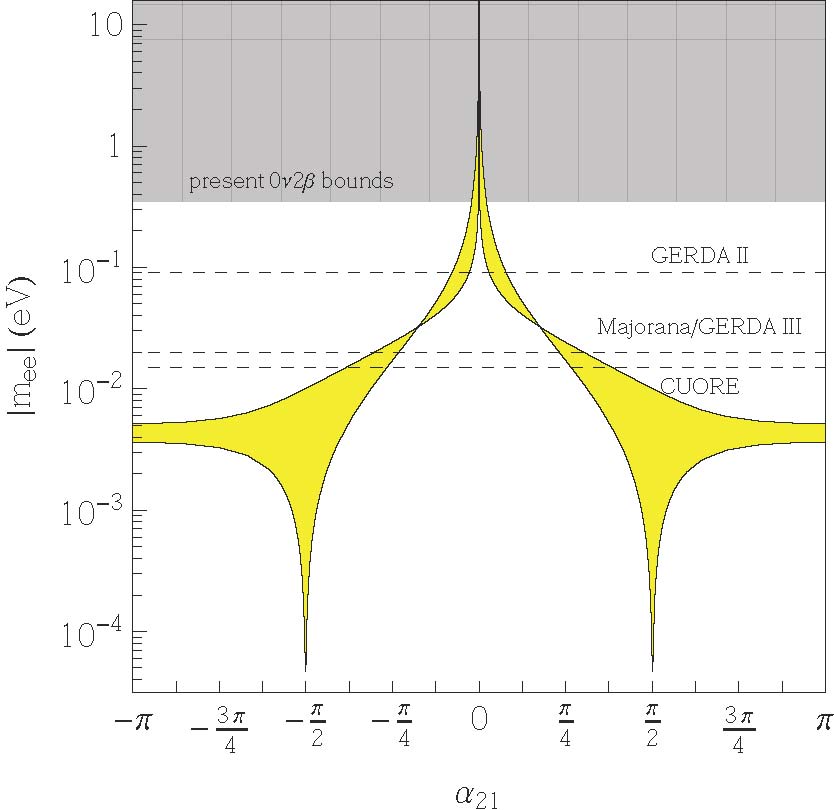}}
 \caption{On the left, $|m_{ee}|$ as a function of the lightest neutrino mass is plotted in the Niemeyer model. The light colored areas represent the possible regions considering only the exact TB pattern: in red the NH and in blue the IH. The purple band is the prediction of the model. Only the NH and the QD can be accommodated. The experimental bounds are the same as in figure \ref{fig1}. On the right, $|m_{ee}|$ as a function of the Majorana phase is plotted. The experimental bounds are the same as in figure \ref{fig3}}
 \label{fig6}
\end{figure}

In figure \ref{fig6}(a), we show $|m_{ee}|$ as a function of the lightest neutrino mass. We see that both the NH and the QD spectrum can be accommodated and furthermore the model does not present a lower bound for the effective mass. From figure \ref{fig6}(b), we see that, within the $3\sigma$ error range for $r$, $|m_{ee}|$ can be zero for a particular value of the Majorana phase.
This plot has been called ``Niemeyer'' plot by the authors due to its similarity to the columns of ``Palacio da Alvorada", designed by the homonym architect, and we refer to this proposal with the name of Niemeyer model.

\begin{figure}[h!]
 \centering
 \subfigure[Comparison between the SS1 and the Niemeyer models.]
   {\includegraphics[width=8cm]{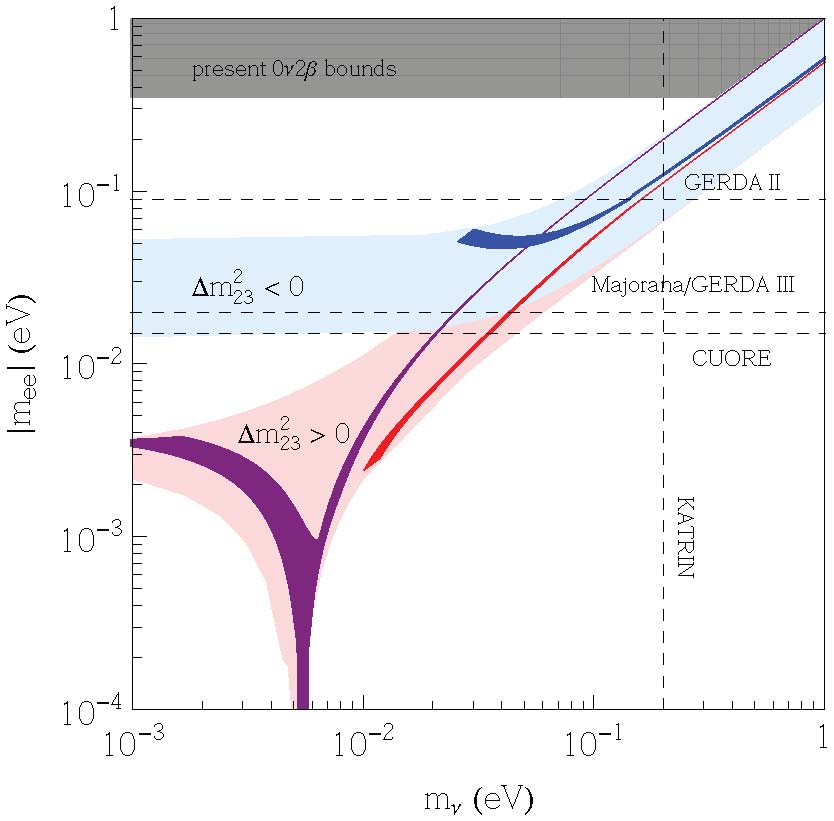}}
 \subfigure[Comparison between the EF and the Niemeyer models.]
   {\includegraphics[width=8cm]{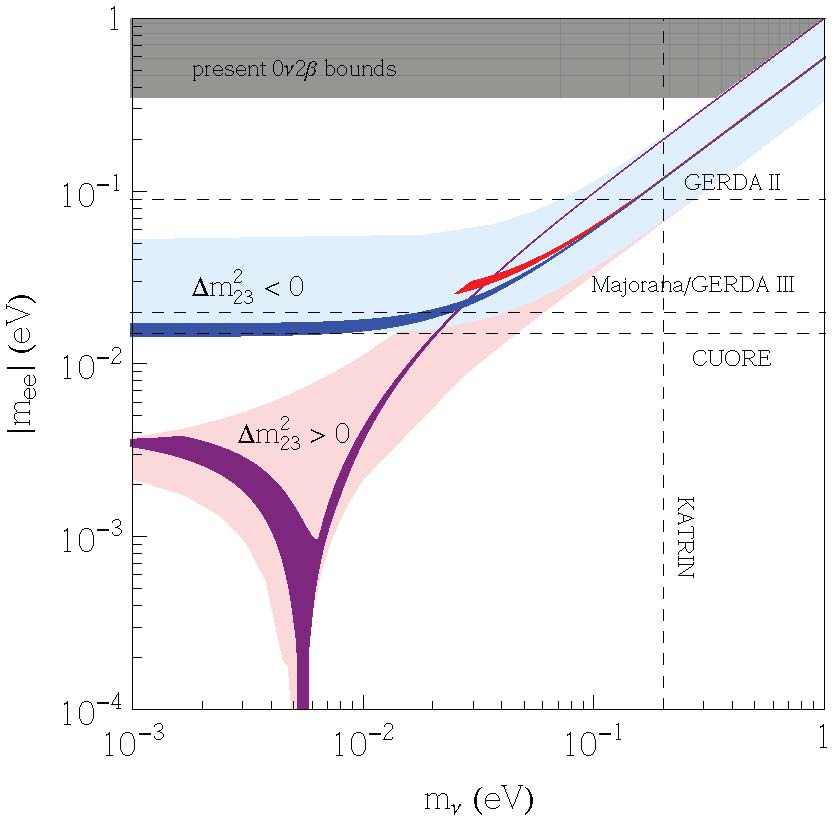}}
 \caption{In both figures, $|m_{ee}|$ as a function of the lightest neutrino mass is plotted. We compare our proposals and the Niemeyer model: on the left we plot the SS1 model and on the right the EF one. In red (green) and blue we plot the NH and the IH, respectively, of our proposals (Niemeyer model). The experimental bounds are the same as in figure \ref{fig1}.}
 \label{fig7}
\end{figure}

\begin{table}[h]
\centering
\begin{tabular}{|c|c|c|c|}
  \hline
  &&&\\[-3mm]
  & $|m_{ee}|$ (meV) & lightest $m_\nu$ (meV) & $\sum_i|m_{\nu i}|$ (meV)\\[3mm]
  \hline
  &&&\\[-3mm]
  Niemeyer model & --- & 0.54 & 58.8 \\[3mm]
  \hline
\end{tabular}
\caption{\label{table:comparisonHMV}Lower bounds for $|m_{ee}|$, the lightest neutrino mass and the sum of the neutrino masses for the Niemeyer model.}
\end{table}

In table \ref{table:comparisonHMV} we present the lower bounds for the lightest neutrino mass and the sum of the neutrino masses for the Niemeyer model and, in figure \ref{fig7}, we compare it with our proposals. The different realizations can be distinguished in two regions of the space of the parameters: in the QD band, the models dispose along two distinct and parallel lines and a combined measure of $|m_{ee}|$ and of the lightest neutrino mass, possible in the next future experiments, can definitively rule out one of the two descriptions; looking at the NH region, in the SS1 and in the EF models there are lower bounds for $|m_{ee}|$ and $m_{\nu_1}$, while in the Niemeyer model they are unconstrained and only when these energy scales will become accessible in the experiments a hope to distinguish among the $A_4$ and the $S_4$ based descriptions will be cherished.\\

It is interesting to note that both the AF and the Niemeyer models, that are based on the $A_4$ symmetry group, present exactly the same behaviour in the quasi degenerate region, which is different from all our proposals, based on the $S_4$ symmetry group.
Furthermore, also the $T'$ model behaves like the AF model and therefore it is not only the kind of the discrete group which discriminates the two profiles. The origin of this splitting between the two curves is the use of the doublet representation: only in the $S_4$ based models a doublet flavon has been introduced in the neutrino sector, in fact in the $T'$ model, the only other one with doublet representations, the doublets are used in the quark sector. Furthermore, in the $T'$ model, it seems to be a very difficult task to reproduce the TB pattern as the lepton mixing matrix introducing a doublet flavon in the neutrino sector.
As a result, a combined measure of the lightest neutrino mass and $|m_{ee}|$ could discriminate between specific flavour models and, moreover, could suggest some particular setting for the scalar sector.\\

\begin{table}[h!]
\centering
\begin{tabular}{|l||cc|cc|c|}
\hline
&&&&&\\[-3mm]
 &SS1& EF& AF See-Saw& AF Effective& Niemeyer\\[3mm]
\hline
&&&&&\\[-3mm]
Normal Hierarchy & X & -- & X & X & X \\[3mm]
&&&&&\\[-3mm]
Inverted Hierarchy & --  & X & X & -- & -- \\[3mm]
\hline
\end{tabular}
\caption{\label{table:comparisonHierarchies}Summary of the type of hierarchies, which can be explained by the SS1, EF, AF with See-Saw case, AF with effective case and Niemeyer models.}
\end{table}

In table \ref{table:comparisonHierarchies}, we summarize the allowed spectrum for all the analyzed models: we have reported only the cases in which the two hierarchies are not in the QD spectrum, because, otherwise, they cannot be easily distinguished by experiments.

%
%

\section{NLO Corrections}
\label{sec:NLO}

We now study the deviations to the LO results of the SS1 model. We first present the analysis for the VEV alignment and then we move to the mass matrices.

\subsection{The VEV Alignment}
Here we only report the results for the vacuum alignment. We consider the higher order operators to the superpotential which leads to the LO alignment for the VEVs and we find that they introduce several corrections. Denoting the general flavon field with $\Phi$, we can write the new VEVs as
\beq
\mean{\Phi_i}=\mean{\Phi_i}^{(LO)}+\de\Phi_i\;,
\eeq
where $\de\Phi_i$ are suppressed by a factor $u$ with respect to $\mean{\Phi_i}^{(LO)}$.
All the perturbations are non vanishing, a part from $\de\eta_1$ and $\de\eta_2$ and one of the perturbations in the neutrino sector, which remains undetermined. On the other hand the NLO terms fixes the relation between $v_\vphi$ and $v_\eta$.

\subsection{The Mass Matrices}
In this part we present the corrections to the mass matrices due to the higher order terms in the matter superpotential and the deviations to the VEV alignment.

The superpotential for the charged leptons can be written as
\beq
w_\ell=w_\ell^0+\de w_\ell
\eeq
where $w_\ell^0$ corresponds to eq.(\ref{eq:wd:leptons}) and $\de w_\ell$ contains all the NLO terms. We note that the LO operators related to $e^c$ completely fill in the first line of $m_\ell$ and, as a result, the corrections can be reabsorbed in the LO parameters. For this reason, we avoid to specify the NLO operators of $\de w_\ell$ related to $e^c$, reporting only those ones connected to $\mu^c$ and $\tau^c$:
denoting $\De$ and $\vphi$ with $\Phi_\nu$ and $\psi$ and $\eta$ with $\Phi_\ell$, we can write
\beq
\dfrac{\tau^c}{\La^2}(\ell\Phi_\ell\Phi_\ell\Phi_\nu+\ell\Phi_\ell\xi'\xi')\;,\qquad\qquad
\dfrac{\mu^c}{\La^3}(\ell\Phi_\nu\Phi_\nu\Phi_\nu+\ell\Phi_\ell\Phi_\ell\xi')\;.
\eeq
These corrections have to be added to those ones originated by $w_\ell^0$ considering the deviations at the NLO to the vacuum alignment. Finally the corrected charged lepton mass matrix has the following structure
\beq
m_\ell=\left(
         \begin{array}{ccc}
           O(u^2t) & O(u^2t) & O(u^2t) \\
           O(u^2) & O(u) & O(u^2) \\
           O(u) & O(u) & O(1) \\
         \end{array}
       \right)uv_d\;,
\eeq
where only the order of magnitude of the single entries is reported. As a consequence the unitary matrix $U_\ell$, which is the transformation of the charged leptons used to diagonalized $m_\ell$, is modified in the following way:
\beq
U_\ell=\left(
         \begin{array}{ccc}
           1 & T_{12}^e u & T_{13}^e u \\
           -T_{12}^e u & 1 & T_{23}^e u \\
           -T_{13}^e u & -T_{23}^e u & 1 \\
         \end{array}
       \right)\;,
\label{eq:NLO:charged leptons}
\eeq
where $T_{ij}^e$ are order one coefficients. We underline that the mass eigenvalues are corrected by terms of relative order $u$ and therefore the LO mass hierarchy is unaffected by the higher order corrections.\\

A similar analysis can be performed for the neutrino superpotential. We first discuss the Dirac neutrino mass matrix $m_\nu^D$. It is easy to see that the first corrections to the first element on eq.(\ref{eq:wd:neutrinos}) appear at the NNLO and the new operators are the following:
\beq
\dfrac{1}{\La^2}(\nu^c\ell\psi\De)h_u\quad\quad\dfrac{1}{\La^2}(\nu^c\ell\psi\vphi)h_u\quad\quad
\dfrac{1}{\La^2}(\nu^c\ell\eta\De)h_u\quad\quad\dfrac{1}{\La^2}(\nu^c\ell\eta\vphi)h_u\;,
\eeq
where the terms in the brackets represent all the possible products, considering the $S_4$ multiplication rules. As a result, we can consider the corrections to the various entries as independent.

For the heavy right-handed neutrino mass matrix $M_N$, there are two sources of corrections: the first corresponds to the insertion of the NLO deviations to the VEVs in the LO superpotential of eq.(\ref{eq:wd:neutrinos}) and they correct each entry with factors of relative order $u$ with a particular scheme, which, however, cannot be reabsorbed in the LO parameters; the second consists in the higher order operators in the superpotential and, at the NLO, the only non vanishing one is
\beq
\dfrac{x_s}{\La}(\nu^c\nu^c\vphi)'\xi'\;,
\eeq
where $x_s$ is an order one coupling constant.

It appears useful to summarize the results of this analysis in this notation for the neutrino mass matrices:
\beq
m_\nu^D=(m_\nu^D)_{LO}+\de m_\nu^D\;u^2\qquad M_N=(M_N)_{LO}+\de M_N\;u
\eeq
where $(m_\nu^D)_{LO}$ and $(M_N)_{LO}$ are those of eq.(\ref{matrices:nu}), $\de m_\nu^D$ is a $3\times3$ matrix of independent order one terms and
\beq
\de M_N=\left(
          \begin{array}{ccc}
            2c\;\frac{\de\De_1}{v_\De u} & b\;\frac{\de\vphi_1}{v_\vphi u}-c\;\frac{\de\De_3}{v_\De u}-a\;\frac{v_\xi'}{\La u} & b\;\frac{\de\vphi_2}{v_\vphi u}-c\;\frac{\de\De_2}{v_\De u}+a\;\frac{v_\xi'}{\La u} \\[1mm]
            b\;\frac{\de\vphi_1}{v_\vphi u}-c\;\frac{\de\De_3}{v_\De u}-a\;\frac{v_\xi'}{\La u} & b\;\frac{\de\vphi_2}{v_\vphi u}+2c\;\frac{\de\De_2}{v_\De u}+a\;\frac{v_\xi'}{\La u} & -c\;\frac{\de\De_1}{v_\De u} \\[1mm]
            b\;\frac{\de\vphi_2}{v_\vphi u}-c\;\frac{\de\De_2}{v_\De u}+a\;\frac{v_\xi'}{\La u} & -c\;\frac{\de\De_1}{v_\De u} & b\;\frac{\de\vphi_1}{v_\vphi u}+2c\;\frac{\de\De_3}{v_\De u}-a\;\frac{v_\xi'}{\La u} \\
          \end{array}
        \right)\,,
\eeq
where $a=2x_sv_\vphi$, $b=2x_dv_\vphi$ and $c=2x_tv_\Delta$.\footnote{Note that with this notation the order of magnitude of the entries of the two matrices $(M_N)_{LO}$ and $\de M_N$ is the same: it is sufficient to remember that $\de\De_i$ and $\de\vphi_i$ are suppressed by a factor $u$ with respect with $v_\De$ and $v_\vphi$, respectively.}

We can now consider the neutrino mass matrix $m_\nu$: looking at eq.(\ref{matrices:LHnu}), we can see that the first corrections come from $\de M_N$. As a result the TB pattern has to be modified in order to diagonalize $m_\nu$ and we can write
\beq
U_\nu=U_{TB}+\de U_\nu\; u
\eeq
where $\de U_\nu$ can be parameterized by three angles, $T_{12}^\nu$, $T_{23}^\nu$ and $T_{13}^\nu$, in a similar way as in eq.(\ref{eq:NLO:charged leptons}).

Finally, summarizing all the corrections from the higher order terms, deviations to the neutrino mixing matrix of relative order $u$ with respect the LO results are generated. The corrected neutrino mixing angles are modified as follows:
\bea
&&\tan\theta_{23}=-1-2u\left(T_{23}^e+\dfrac{\sqrt2T_{13}^\nu-2T_{23}^\nu}{\sqrt3}\right)\\[0.3cm]
&&\tan\theta_{12}=\dfrac{1}{\sqrt2}-\dfrac{3u}{4}\left(\sqrt2\left(T_{12}^e+ T_{13}^e\right)-\sqrt3\left(T_{12}^\nu+T_{13}^\nu\right)\right)\\[0.3cm]
&&\tan\theta_{13}=\dfrac{u}{2\sqrt3}\left(\sqrt6\left(T_{12}^e-T_{13}^e\right)+T_{13}^\nu+2\sqrt2T_{23}^\nu-3T_{12}^\nu\right)\;.
\eea
We can conclude that the NLO corrections originate deviations to the TB mixing angles of order $u$.

%
%

\section{Type II and III See-Saw Mechanisms}
\label{sec:typeIIandIII}

In the previous sections we have analyzed in details the SS1 model. However it represents only one possible manner of generating neutrino masses. The different possibilities at our disposal can be divided into two classes, characterized by inducing neutrino masses radiatively or not. In the present paper we concentrate on the latter class of models, of which SS1 is the best known example. As fixed points in our analysis we require that the different realizations present the same flavour group, $G_f= S_4\times Z_5\times U(1)_{FN}$, the same flavons, and the same number of free parameters to describe the structure of the neutrino mass matrix as in the SS1 model. For these reasons, we discard the combinations of different mechanisms, which would bring more parameters than the traditional ones.  Furthermore, these conditions define a class of equivalent models, whose phenomenologies and predictions can be compared.

\subsection{Type II See-Saw}
\label{sec:sub:TypeII}

With type II See-Saw (SS2) mechanism we refer to the general scenario in which neutrinos get a mass thanks to the coupling of the SU(2) lepton doublet $\ell$ with a $SU(2)$ heavy Higgs triplet of hypercharge $+1$. Neutrino masses are generated when the neutral component of the scalar triplet develops a small VEV.\footnote{Alternatively we can integrate out the heavy Higgs triplet. In this case we get an effective five-dimensional operator $\sim \lambda  \ell \ell h_u h_u/ M_\Upsilon$ where $M_\Upsilon$ is the mass of the heavy Higgs triplet.  The two picture are phenomenologically identical since the minimization of the potential leads  to $v_\Upsilon = \lambda h_u h_u/M_\Upsilon$, as written in the text.} With respect to what presented in \cite{BMM}, the implementation of this model is straightforwardly realized by adding to the field scalar content a couple of $SU(2)$ Higgs triplets $\Upsilon,\bar{\Upsilon}$ of hypercharge $+1$ and $-1$ respectively and neutral under  $G_f$. As usual the presence of $\bar{\Upsilon}$ is necessary both to cancel anomalies and to introduce a mass term for $\Upsilon$ in the superpotential. \footnote{It is relevant to clarify that $\bar{\Upsilon}$ is a chiral superfield, which works as the complex conjugate of $\Upsilon$: this requirement is motivated by the supersymmetry context in which we are working, but in a non-supersymmetric one $\bar{\Upsilon}$ would be just the complex conjugate of $\Upsilon$.}

\begin{table}[h]
\begin{center}
\begin{tabular}{|c||c|c|c|c|c|c||c||c|c||c|c||c|}
  \hline
  &&&&&&&&&&&& \\[-0,3cm]
  & $\ell$ & $e^c$ & $\mu^c$ & $\tau^c$ & $h_{u,d}$&$\Upsilon,\bar{\Upsilon}$ & $\theta$ & $\psi$ & $\eta$ & $\Delta$ & $\vphi$ & $\xi'$ \\
  &&&&&&&&&&&& \\[-0,3cm]
  \hline
  &&&&&&&&&&&& \\[-0,3cm]
  $S_4$ & $3_1$ & $1_2$ & $1_2$ & $1_1$ & $1_1$ &$1_1$& $1_1$ & $3_1$ & 2 & $3_1$ & $2$ & $1_2$  \\
  &&&&&&&&&&&& \\[-0,3cm]
  $Z_5$ & $\om$ & $\om^3$ & 1 & $\om^2$ & 1 &1& 1 & $\om^2$ & $\om^2$ & $\om^3$ & $\om^3$ & 1  \\
  &&&&&&&&&&&& \\[-0,3cm]
  $U(1)_{FN}$ & 0 & 1 & 0 & 0 & 0&0 & -1 & 0 & 0 & 0 & 0 & 0  \\
  \hline
  \end{tabular}
\end{center}
\begin{center}
\begin{minipage}[t]{12cm}
\caption{Transformation properties of the lepton and scalar   fields  and of all the flavons of the model.  $\Upsilon,\bar{\Upsilon}$ are $SU(2)$ scalar triplets with opposite  hypercharge.}
\label{table:lepton_transformationTypeII}
\end{minipage}
\end{center}
\end{table}

The superpotential for the charged leptons is the same as in eq.(\ref{eq:wd:leptons}), while that one for neutrinos becomes
\bea
w_\nu\;=&& \dfrac{x_d}{\La}(\ell \ell \Upsilon \vphi)+\dfrac{x_t}{\La}(\ell \ell \Upsilon \Delta)+\hc+\ldots
\label{eq:wd:neutrinos-II}
\eea
When $\Upsilon$ and $\bar{\Upsilon}$ get a non-vanishing VEV
\beq
\mean{\Upsilon}\,\equiv\, v_\Upsilon\,\equiv\,\mean{\bar{\Upsilon}}
\eeq
and the flavons get the VEVs reported in eq.(\ref{vev:allleptons}), the neutrino mass matrix has the same structure of $M_N$ in eq.(\ref{matrices:nu}), but it is multiplied by the additional factor $v_\Upsilon$,
\beq
m_\nu=\left(
              \begin{array}{ccc}
                2c & b-c & b-c \\
                b-c & b+2c & -c \\
                b-c & -c & b+2c \\
              \end{array}
            \right) \,v_\Upsilon\;,
\eeq
where the parameters $b$ and $c$ are now equal to $2x_d v_\vphi/\Lambda$ and $2x_t v_\Delta/\Lambda$, respectively. Due to the similarity to $M_N$ in eq.(\ref{matrices:nu}), this light neutrino mass matrix is diagonalizable by the TB pattern.

Before commenting on the phenomenology of this realization, we briefly remember the mass matrix for the neutrinos in the EF model, which is
\beq
m_\nu^{EF}=\left(
              \begin{array}{ccc}
                2c & b-c & b-c \\
                b-c & b+2c & -c \\
                b-c & -c & b+2c \\
              \end{array}
            \right)\dfrac{v_u^2}{\Lambda}\;,
\eeq
where $b$ and $c$ are the same as in the previous equation. If we identify $v_\Upsilon$ with $v_u^2/\Lambda$, the two matrices coincide and then the phenomenological analysis performed for the EF model is valid for the SS2 model. The discrimination between these two scenarios can be done only by considering the Higgs scalar sector. A first comment refers to the order of magnitude of $v_\Upsilon$, which has to be relatively small in order to explain the lightness of the neutrinos: with a naively analysis, considering $|\De m_{atm}^2|^{1/2}$ as the typical energy scale for the neutrino masses and that the parameters $b$ and $c$ are of order $u$, we get that $v_\Upsilon$ has to be of order of few eV. The VEV of the Higgs triplet is usually induced by the VEV of the Higgs doublet $h_u$: the superpotential presents the terms
\begin{equation}
M_\Upsilon \Upsilon \bar{\Upsilon} +\lambda \bar{\Upsilon} h_u h_u\;,
\end{equation}
from which we get the following terms in the scalar potential
\begin{equation}
M^2_\Upsilon |\bar{\Upsilon} |^2+M^2_\Upsilon |{\Upsilon} |^2 +\lambda M_\Upsilon   \Upsilon h_uh_u+\ldots\;.
\end{equation}
Once we include the soft terms and $h_u$ acquires VEV, the induced VEV for $\Upsilon$ is given by
\begin{equation}
v_\Upsilon = \lambda\dfrac{v_u^2}{M_\Upsilon}\,.
\end{equation}
In order to have $v_\Upsilon\sim 1$ eV we need that $\lambda/ M_\Upsilon \sim 10^{-13} \,\mbox{GeV}^{-1}$. On the other hand the best collider signatures for the SS2 model, such as $\Upsilon^{\pm\pm}\to \ell^\pm\ell^\pm$ \cite{MaAccomandoGarayoa} are  accessible for $M_\Upsilon \sim$ TeV. Therefore there is a tension between the naturalness of the supersymmetric SS2 model and its phenomenological interest. We mention that by slightly modifying the model proposed (letting, for example, $h_u$ transform under the $U(1)_{FN}$) in a completely natural way, it is possible to lower the value for $\lambda$ and as consequence to lower the scale of $M_\Upsilon$ down to the TeV range.

\subsection{Type III See-Saw}
\label{sec:sub:TypeIII}

In the type III See-Saw (SS3) mechanism, we add to the SM fermionic content of the EF model three $SU(2)$ fermion triplets $\Sigma$ uncharged under $U(1)_Y$.\footnote{It is also possible to introduce only two such fermion triplets: indeed it is the minimal number of $\Sigma$ in order to fit the data. With three, $\Sigma$, however, it is possible to describe three non-vanishing masses.} These triplets present a Majorana mass term and  couple through a trilinear term to the SU(2) lepton doublet $\ell$ and to the Higgs doublet $h_u$: the coupling provides the neutrino masses and introduces little mixings between the charged leptons. In this realization with the type III See-Saw mechanism, the new fermion triplets behave under $G_f$ as the right-handed neutrinos of the SS1 model presented in section \ref{sec:typeI} and they induce the neutrino masses when electroweak (EW) symmetry is broken.

\begin{table}[h]
\begin{center}
\begin{tabular}{|c||c|c|c|c|c|c||c||c|c||c|c||c|}
  \hline
  &&&&&&&&&&&& \\[-0,3cm]
  & $\ell$ & $e^c$ & $\mu^c$ & $\tau^c$ & $\Sigma$ & $h_{u,d}$ & $\theta$ & $\psi$ & $\eta$ & $\Delta$ & $\vphi$ & $\xi'$ \\
  &&&&&&&&&&&& \\[-0,3cm]
  \hline
  &&&&&&&&&&&& \\[-0,3cm]
  $S_4$ & $3_1$ & $1_2$ & $1_2$ & $1_1$ & $3_1$ & $1_1$ & $1_1$ & $3_1$ & 2 & $3_1$ & $2$ & $1_2$  \\
  &&&&&&&&&&&& \\[-0,3cm]
  $Z_5$ & $\om^4$ & $1$ & $\om^2$ & $\om^4$ & $\om$ & 1 & 1 & $\om^2$ & $\om^2$ & $\om^3$ & $\om^3$ & 1  \\
  &&&&&&&&&&&& \\[-0,3cm]
  $U(1)_{FN}$ & 0 & 1 & 0 & 0 & 0 & 0 & -1 & 0 & 0 & 0 & 0 & 0  \\
  \hline
  \end{tabular}
\end{center}
\begin{center}
\begin{minipage}[t]{12cm}
\caption{Transformation properties of the lepton and scalar fields and of all the flavons of the model. $\Sigma$ are $SU(2)$ fermion triplets with vanishing hypercharge.}
\label{table:lepton_transformationTypeIII}
\end{minipage}
\end{center}
\end{table}

The superpotential of the model in the lepton sector is given by the sum of three pieces:
\beq
w_\ell+w_\nu+w_\Sigma\;,
\eeq
where
\bea
w_\ell\;=&&\sum_{i=1}^{4}\dfrac{\theta}{\La}\dfrac{y_{e,i}}{\La^3}e^c(\ell X_i)'h_d+\dfrac{y_\mu}{\La^2}\mu^c(\ell\psi\eta)'h_d+\dfrac{y_\tau}{\La}\tau^c(\ell\psi)h_d+\hc+\ldots\\[0.3cm]
w_\nu\;=&&x(\Sigma \ell)h_u+\hc+\ldots\\[0.3cm]
w_\Sigma\;=&&x_d \, \Sigma \Sigma \vphi + x_t \Sigma \Sigma \Delta+\hc+\ldots\,.
\eea
and $X_i$ is the same of eq.(\ref{eq:X}).
$\Sigma$ has three components defined as $\Sigma^a= s^a\sigma^a$, where $\sigma^a$ are the Pauli matrices, generators of $SU(2)$, and the Yukawa term $(\Sigma \ell)h_u$ understands the $SU(2)$ gauge invariant contractions $ \sigma^a_{\rho \tau}\Sigma^a \ell_\rho (h_u)_\tau$.

When the flavour and the EW symmetries are broken, both charged leptons and neutrinos mix with the $\Sigma$ components. Assuming that the flavons get VEVs as reported in eq.(\ref{vev:allleptons}), for the neutrinos we have
\bea
M_{\nu} &=& \left(\begin{array}{cc} 0 & (m_\nu^D)^T \\m_\nu^D& M_\Sigma \end{array} \right)\,,
\eea
with $m_{\nu}^D$ and $M_\Sigma$ are equal to $m_{\nu}^D$ and $M_N$ in eq.(\ref{matrices:nu}),
\beq
m_{\nu}^D=\left(
          \begin{array}{ccc}
            1 & 0 & 0 \\
            0 & 0 & 1 \\
            0 & 1 & 0 \\
          \end{array}
        \right)xv_u\qquad\qquad
M_\Sigma=\left(
          \begin{array}{ccc}
            2c & b-c & b-c \\
            b-c & b+2c & -c \\
            b-c & -c & b+2c \\
          \end{array}
        \right)\;,
\label{eq:mSig}
\eeq
where $b=2x_dv_\vphi$ and $c=2x_tv_\Delta$. It appears evident now that $\Sigma$ take the roles of $\nu^c$ of the SS1 model. Moreover as $M_N$, $M_\Sigma$ is exactly diagonalized by the TB mixing and therefore
\beq
M_\Sigma^{diag}=(3c-b,\;2b,\;3c+b)\;.
\eeq
Integrating out the heavy degrees of freedom, we get the light neutrino mass matrix, which is given by the usual relation
\beq
m_\nu=-(m_\nu^D)^TM_\Sigma^{-1}m_\nu^D\;,
\label{matrices:LHnuTypeIII}
\eeq
still diagonalizable by the TB pattern. As a result the neutrino mass eigenvalues can be written in the same way as in eq.(\ref{LNMassEigen}) and it is possible to estimate the value of the mass of the fermion triplet $\Sigma$ using the expression in eq.(\ref{stimaMN}) only substituting $M_\Sigma$ to $M_N$:
\beq
m_\nu\sim\dfrac{x^2v_u^2}{M_\Sigma}
\label{stimaMS}
\eeq
and, taking again $v_u=174$ GeV, $x$ of $\mcal{O}(1)$ and $|\De m^2_{atm}|^{1/2}$ as the typical light neutrino mass scale, we get a typical value for $M_\Sigma$ of about $5\times10^{14}$ GeV.

It is clear that for what concerns the phenomenology of the light neutrinos, the SS3 model cannot be distinguished from SS1 one. The two descriptions may be discriminated by taking into account phenomenological processes involving the heavy neutral fermions, such as leptogenesis (see section \ref{sec:Leptogenesi}), or the charged lepton sector that presents  a small mixing  with the charged components of $\Sigma$.
The full mass matrix for the charged sector is given by ($M_\ell\sim RL$)
\bea
M_\ell &=& \left(\begin{array}{cc} m_\ell & 0\\
m_{\ell\Sigma} & M_\Sigma \end{array} \right)\,,
\eea
with $m_\ell$ identical to the mass matrix in eq.(\ref{matrices:l}), $m_{\ell\Sigma}\equiv m_\nu^D$ and $M_\Sigma$ reported above. The perturbations induced by $\Sigma$ to $m_{\ell}$ are completely negligible since they are of $\mathcal{ O}(m_{\ell\Sigma}/M_\Sigma)\ll u$ and as a result the TB predictions for the lepton mixing are unchanged.
However, the presence of the triplet induces LFV decays such as $\mu\to e \gamma$, $\tau\to e\gamma$ and  $\tau\to\mu\gamma$ and the results presented in \cite{Abada} are valid also in the context of the model so far proposed.

%
%

\section{Leptogenesis}
\label{sec:Leptogenesi}

In this section we deal with the constraints on our models from leptogenesis (see \cite{Davidson} and references therein). We point at getting some constraints on the parameter $u$, imposing that the lepton asymmetry parameter $\epsilon$ is large enough in order to provide an adequate baryon asymmetry.
We start considering the SS1 model, which is the most discussed in literature, and after we approach the situations for the SS2 and SS3 models.\\

\subsection{SS1 Model}
The out-of-equilibrium decays of the right-handed neutrinos $\nu^c$ in the early universe to lepton and Higgs doublets produce lepton asymmetries. Since the masses of the right-handed neutrinos are very heavy, with masses above $10^{13}$ GeV, we assume the unflavoured framework.

In the basis in which $M_N$ is diagonal and real, the lepton asymmetry parameters (in the supersymmetric context) are expressed through the following relation \cite{Davidson}
\beq
\epsilon_i=\dfrac{1}{2\pi\left(Y_\nu Y_\nu^\dag\right)_{ii}}\sum_{j\neq i}\rm{Im}\left\{\left[\left(Y_\nu Y_\nu^\dag\right)_{ij}\right]^2\right\}f\left(\dfrac{|M_j|^2}{|M_i|^2}\right)
\label{epsilon}
\eeq
where $Y_\nu\equiv m_\nu^D/v_u$. The off-diagonal entries of the term $Y_\nu Y_\nu^\dag$ are the dominant contributions on $\epsilon_i$: at LO we see from eq.(\ref{matrices:nu}) that
\beq
Y_\nu Y_\nu^\dag=|y|^2\unity\;.
\eeq
In order to go in the basis in which $M_N$ is diagonal, we have to apply to $Y_\nu$ the unitary matrix $U_{TB}$ and as a result
\beq
Y_\nu\longrightarrow U_{TB}^T Y_\nu\;;
\eeq
even if this rotation, the product $Y_\nu Y_\nu^\dag$ remains proportional to the unity matrix. We can conclude that $\epsilon_i$ vanishes at LO.

From the analysis of the NLO corrections, we get that the first correction to $Y_\nu$ is of order $u^2$ and this is also the order of the first off-diagonal entries of the product $Y_\nu Y_\nu^\dag$:
\beq
Y_\nu Y_\nu^\dag\sim (Y_\nu)_{LO}(Y_\nu)_{LO}^T+(Y_\nu)_{LO}\de Y_\nu^T\,u^2+\de Y_\nu(Y_\nu)_{LO}^T\,u^2\;.
\label{YY+}
\eeq
Moving to the basis in which $M_N$ is diagonal, this leaves unchanged the order of magnitude of the off-diagonal entries of $Y_\nu Y_\nu^\dag$: the first term in the right-handed part of eq.(\ref{YY+}) is the unity matrix; the second and the third terms are affected by the rotations of the TB pattern, but the corrections $\de U_\nu$ appear at higher orders with respect with those ones shown in eq.(\ref{YY+}).

Before discussing the lepton asymmetry parameters $\epsilon_i$, we observe that the contribution from the function $f\left(|M_j|^2/|M_i|^2\right)$ is not trivial: in our model the right-handed neutrinos are strongly hierarchical and therefore we can use the following expression for $f$
\beq
f\left(\dfrac{|M_j|^2}{|M_i|^2}\gg1\right)=-3\sqrt{\dfrac{|M_i|^2}{|M_j|^2}}\;.
\label{fApprox}
\eeq
For the NH case, the lightest right-handed neutrino is $M_3$ and leptogenesis is governed by $\epsilon_3$, namely by $\left(Y_\nu Y_\nu^\dag\right)_{3i}$, $i=1,2$, which are non zero and complex. Combining the two relations in eq.(\ref{epsilon}) and in eq.(\ref{fApprox}), we can finally write the following expression for $\epsilon_3$
\beq
|\epsilon_3|=\sum_{i=1,2}\dfrac{3}{2\pi}\dfrac{|M_3|}{|M_i|}\,\mcal{O}(u^4)\,.
\eeq
Considering that $|M_{1,2}|\sim4|M_3|$ and that the requirement for the model to produce a lepton asymmetry sufficient to explain an adequate baryon asymmetry corresponds to $|\epsilon_3|\gtrsim10^{-7}$, we get a lower bound on $u$ of about $0.02$, which is in the range of values, eq.(\ref{vev:uet}), required in order to explain the correct fermion mass hierarchies and mixings.

\subsection{SS2 Model}

In this part we discuss the case of the type II See-Saw model. Since the mechanism involves lepton number violation and allows for new CP violating processes, it is interesting to examine its relationship with leptogenesis \cite{LeptoIISM}. Let give a glance on the non-supersymmetric context, in which $\bar{\Upsilon}$ is the complex conjugate of $\Upsilon$. The CP asymmetry that is induced by the triplet scalar decays is defined as
\beq
\epsilon_\Upsilon\equiv2\dfrac{\Gamma\left(\bar{\Upsilon}\rightarrow\ell\ell\right)- \Gamma\left(\Upsilon\rightarrow\bar{\ell}\bar{\ell}\right)}{\Gamma_\Upsilon+\Gamma_{\bar\Upsilon}}\;,
\eeq
where the overall factor 2 remember the presence of two (anti)leptons. While a single triplet is enough to produce three light massive neutrinos, the lepton number asymmetry is generated only at higher loops and it results unacceptably small. The situation improves introducing additional sources for the neutrino masses, such as singlet fermions, triplet fermions or different triplet scalars.

Moving to the supersymmetric framework, in which $\bar{\Upsilon}$ is a distinct superfield from $\Upsilon$, as defined in section \ref{sec:sub:TypeII}, the analysis changes drastically: the so-called soft leptogenesis, which is driven by soft supersymmetry breaking terms, can be successful with only the couple $\Upsilon$-$\bar{\Upsilon}$ in reproducing neutrino masses and the lepton number asymmetry\cite{LeptoIISUSY}. In this case a complete analysis of the soft supersymmetry breaking terms is needed, but this is beyond the aim of this project. However, assuming universal soft trilinear couplings in the range $(0.1-10)$ TeV, the mass of the Higgs triples $M_\Upsilon$ is required to belong to the range $(10^3-10^9)$ GeV. A larger value would lead to a smaller lepton asymmetry, which could not explain the baryon asymmetry. As discussed in section \ref{sec:sub:TypeII}, the natural value for $M_\Upsilon$ is about $10^{13}$ GeV, larger than the favorite range for leptogenesis.

\subsection{SS3 Model}

In this part we discuss the case of the type III See-Saw model. What follows is valid for both the supersymmetric and the non-supersymmetric framework, indeed the final results are modified only by $\mcal{O}(1)$ factors.

The formalism and the qualitative features are very similar to the singlet fermion case for what concerns neutrino masses, as we have seen in section \ref{sec:sub:TypeIII}. Moving to leptogenesis, there are, however, some differences. A significant qualitative new aspect regards the fact that the triplet has gauge interactions. The effect on the washout factor is particularly significant in the so-called "weak washout regime'', i.e. for $\tilde{m}\ll10^{-3}$ eV, where $\tilde{m}$ is defined as an effective mass factor related to the triplet mass $M_\Sigma$ and the diagonal entries of the Dirac neutrino mass matrix, in section (\ref{sec:sub:TypeIII}), by the relation
\beq
\tilde{m}_i\equiv\left|Y_\nu Y_\nu^T\right|_{ii}\dfrac{v_u^2}{M_{\Sigma\,i}}\;.
\eeq
On the other hand, for $\tilde{m}\gg10^{-3}$ eV, the Yukawa interactions are responsible for keeping the heavy fermion abundance close to the thermal equilibrium, so the difference in the washout factor, and therefore in the leptogenesis, between the singlet and the triplet case is only $\mcal{O}(1)$. In our model, the diagonal entries of $Y_\nu$ are all equal and proportional to $x$, which is $\mcal{O}(1)$, in the absence of any particular assumption. As a result $\tilde{m}_3$, which is the largest one, is defined only by the mass of the triplet fermion $M_{\Sigma\,3}$ and therefore $\tilde{m}_3\gg10^{-3}$eV. We can conclude that the triplet leptogenesis shows the same results of the singlet one and then the same constraints on the parameter $u$.

%
%

\section{Conclusions}
\label{sec:Conclusions}

In our previous paper \cite{BMM}, we have proposed a model based on the flavour group $S_4\times Z_5\times U(1)_{FN}$, which is able to reproduce fermion masses and mixings. An effective approach, with the Weinberg operator, has been adopted in order to describe the neutrino masses. In the present paper we have investigated on possible origins for these effective terms and we have constructed some realizations based on the flavour group $S_4\times Z_5\times U(1)_{FN}$, where the neutrino masses are explained through the See-Saw mechanisms. All the solutions we have proposed give essentially two phenomenological scenarios: in the first one, the neutrino spectrum can be only normally hierarchical and $|m_{ee}|$ can reach very low values, at about few meV, while, in the second one, the spectrum can be only of the inverted type and there is a lower bound for $|m_{ee}|$ which is about a order of magnitude higher than in the previous case. In particular the type I and III are similar and refer to the first scenario, while the type II to the second one.

The next future experiments could be able to distinguish among these three realizations: the forthcoming $0\nu2\be$-decay experiments could discriminate between SS1 and SS2 models, but they cannot say too much for the comparison between SS1 and SS3 models. Some differences among these proposals can be found looking at some LFV decays such as $\mu\to e\ga$, $\tau\to e\ga$ and $\tau\to\mu\gamma$, which are affected by the different couplings of the fermion singlets and triplets. Considering that the next MEG experiment \cite{MEG} will improve the sensitivity on $\mu\to e\gamma$, a deepening study on some LFV transitions in the frameworks of the SS1 and SS3 models should be done. Some examples in this direction have already appeared, like for example in the Minimal Flavour Violation (MFV) \cite{MFV} models and in the AF model \cite{fhlm:LFV}.

The introduction of new physics beyond the Standard Model, like right-handed neutrinos, scalar triplets or fermion triplets, could originate a non-vanishing Leptogenesis: indeed in all the See-Saw frameworks it is possible to account for CP violating processes. The SS1 and SS3 models provides quite similar indications for leptogenesis: only taking a value for $u$ of about $0.01\div0.02$ the models can account for a sufficiently large lepton number asymmetry in order to explain the measured baryon asymmetry. Regarding the SS2 model, we conclude that the prediction for the CP violation parameter $\epsilon$ is too small and consequently leptogenesis cannot explain the baryon asymmetry. This result is linked to the masses of the scalar triplets, which are naturally high in these kind of realizations: we underline that lower masses would lead to acceptable leptogenesis and some modifications in this direction can be easily implemented.

Our constructions share the guideline with many other models, proposed in literature. The main task is to get at LO the TB pattern as the neutrino mixing matrix, using a discrete flavour symmetry group $G_f$. Usually some scalar fields are introduced, which get non-vanishing VEVs and spontaneously break $G_f$ down to its subgroup $Z_2$ in the neutrino sector and to nothing or to a different subgroup in the charged lepton one. This breaking chain is fundamental in these kind of models, because the remaining $Z_2$ subgroup represents the low energy flavour structure of the neutrino mass matrix. In our case, $G_f=S_4$ is spontaneously broken down to $Z_2\times Z_2$, but the final neutrino mass matrix is unaffected by the additional $Z_2$ factor. Moreover, further symmetries are added to $S_4$ and in our cases it the product $Z_5\times U(1)_{FN}$. This factor helps in suppressing some unwanted and dangerous terms both in the lepton and in the quark sectors. As a result, the mixings in the charged leptons are negligible and then the lepton mixing matrix coincides with the neutrino one, the TB scheme, where the TB angles are the following
\[
\sin^2\theta_{13}^{TB}=0\qquad\sin^2\theta_{23}^{TB}=1/2\qquad\sin^2\theta_{12}^{TB}=1/3\;.
\]

The next experiments, which will try to measure the $\nu_e$ appearance, like DOUBLE CHOOZ, Daya Bay and MINOS, will improve the sensitivity on the reactor angle $\theta_{13}$ and if a non-vanishing value is found the TB pattern will need some new ingredient. An easy way out to this eventuality is to introduce small corrections to the TB scheme. The higher order terms work in this direction: considering the NLO corrections, the predicted angles are modified by terms of relative order $u$ with respect with the LO ones and, in particular, the reactor angle in not vanishing any more, but it is comparable with $u$.

In spite of these common features, our proposals can be distinguished from other present descriptions, which provide the TB pattern as the lepton mixing matrix: we have considered two phenomenologically interesting and widespread models, the AF and the Niemeyer realizations. A further proposal \cite{fhlm:Tp} bases on the $T'$ group could in principle be considered for the comparison but it provides exactly the same feature in the lepton sector of the AF model.

\begin{table}[h]
\centering
\begin{tabular}{|c|c|c|c|}
  \hline
  &&&\\[-3mm]
  & $|m_{ee}|$ (meV) & lightest $m_\nu$ (meV) & $\sum_i|m_{\nu i}|$ (meV)\\[3mm]
  \hline
  &&&\\[-3mm]
  SS1 and SS3 & 2.3 & 9.5 & 67.3 \\[3mm]
  SS2 and EF & 14.4 & 0.72 & 89.4 \\[3mm]
  AF/$T'$ --- See-Saw case with NH & 5.8 -- 9.2 & 4.4 -- 7.3 & 57.9 -- 73.6 \\[3mm]
  AF/$T'$ --- See-Saw case with IH & 15.2 & 15.7 & 109.6 \\[3mm]
  AF/$T'$ --- Effective case & 3.7 & 13.8 & 77.2 \\[3mm]
  Niemeyer model & --- & 0.54 & 58.8 \\[3mm]
  \hline
\end{tabular}
\caption{\label{table:comparisonConcl}Lower bounds for $|m_{ee}|$, the lightest neutrino mass and the sum of the neutrino masses. The first row refers to the SS1 and SS3 models, the second row to the SS2 one, the third, the fourth and the fifth rows to the AF and $T'$ models with and without See-Saw mechanism, and finally the sixth row to the Niemeyer model. For the AF model with the See-Saw mechanism in the NH case we report the whole ranges for the three observables.}
\end{table}

In table \ref{table:comparisonConcl}, we summarize the different lower bounds on $|m_{ee}|$, on the lightest neutrino mass and on the sum of the masses, achieved by all the models illustrated in the paper. For the AF model with the See-Saw mechanism in the NH case we report the entire ranges for the three observables: this is the unique realization, among those studied in this paper, which presents upper bounds for $|m_{ee}|$, the lightest neutrino mass and the sum of the neutrino masses. Of course, with the improvement of the sensitivity of the next future experiments, some of these proposals can be directly tested and confirmed or ruled out.

Furthermore, we underline the interesting profiles representing $|m_{ee}|$ as a function of the lightest neutrino mass: in the quasi degenerate region, the curves related to the $A_4$ and to the $T'$ based models dispose along the same line; on the other hand the curves related to the $S_4$ based models dispose along a distinct and parallel line with respect to the previous one. The origin of this different behaviour has to be addressed to the use of the doublet representation: only in the $S_4$ based models a doublet flavon has been introduced in the neutrino sector, in fact in the $T'$ model, the only other one with doublet representations, the doublets are used in the quark sector. Furthermore, in the $T'$ model, it seems to be a very difficult task to reproduce the TB pattern as the lepton mixing matrix introducing a doublet flavon in the neutrino sector. As a result, a combined measure of the lightest neutrino mass and $|m_{ee}|$ could discriminate between specific flavour models and, moreover, could suggest some particular setting for the scalar sector.\\

We conclude expressing our trust on the next generation experiments, whose results will be relevant in order to better understand the nature of the physics beyond the Standard Model and hopefully the flavour symmetry, which could describe neutrino oscillations as well as mass hierarchies.

%
%

\section*{Acknowledgments}
We thank Martin Hirsch and Yin Lin for useful comments and discussions.
The work of FB has been partially supported by MEC-Valencia  MEC grant FPA2008-00319/FPA, by European Commission Contracts
MRTN-CT-2004-503369 and ILIAS/N6 RII3-CT-2004-506222 and by the foundation for Fundamental Research of Matter (FOM) and the National Organization for Scientific Research (NWO). LM recognizes that this work has been partly supported by the European Commission under contract MRTN-CT-2006-035505.
The work of SM supported by MEC-Valencia  MEC grant FPA2008-00319/FPA, by European Commission Contracts
MRTN-CT-2004-503369 and ILIAS/N6 RII3-CT-2004-506222.

%
%

\end{document}